\documentclass[12pt]{article}
\usepackage{euscript,amsmath, amssymb, amsfonts}

\newcommand{ \supp}{{\rm supp\,}}
\newcommand{\R}{{\mathbb R}}

\newcommand{\oC}{{\mathbb C}}
\newcommand{\oP}{{\cal P}}
\newcommand{\di}{{ d }}
\newcommand{\D}{\EuScript D}
\newcommand{\eE}{\EuScript E}
\newcommand{\DD}{\overleftarrow D}
\newcommand{\G}{{\mathbb G}}
\newcommand{\Z}{{\mathbb Z}}
\newcommand{\K}{{\mathbb K}}
\newcommand{\be}{\begin{equation}}
\newcommand{\ee}{\end{equation}}

\newcounter{subsubsub}

\newcounter{theorem}
\newcommand{\theorem}{\par\refstepcounter{theorem}
{\bf Theorem \arabic{section}.\arabic{theorem}. }}
\renewcommand\thetheorem{\thesection.\arabic{theorem}}
\makeatletter \@addtoreset{theorem}{section}

\newcounter{lemma}
\newcommand{\lemma}{\par\refstepcounter{lemma}
{\bf Lemma \arabic{section}.\arabic{lemma}. }}
\renewcommand\thelemma{\thesection.\arabic{lemma}}
\makeatletter \@addtoreset{lemma}{section}

\newcounter{proposition}
\newcommand{\proposition}{\par\refstepcounter{proposition}
{\bf Proposition \arabic{section}.\arabic{proposition}. }}
\renewcommand\theproposition{\thesection.\arabic{proposition}}
\makeatletter \@addtoreset{proposition}{section}

\newcounter{definition}
\newcommand{\definition}{\par\refstepcounter{definition}
{\bf Definition \arabic{section}.\arabic{definition}. }}

\makeatletter \@addtoreset{definition}{section}

\font\frtnfr = eufm10 scaled\magstep1
\font\twlfr = eufm10
\font\tenfr = eufm10

\newfam\frfam
\textfont\frfam = \frtnfr
\scriptfont\frfam = \twlfr
\scriptscriptfont\frfam = \tenfr

\font\frtnopen = msbm10 scaled\magstep2
\font\twlopen = msbm10
\font\tenopen = msbm10

\newfam\openfam
\textfont\openfam = \frtnopen
\scriptfont\openfam = \twlopen
\scriptscriptfont\openfam = \tenopen
\def\open{\fam\openfam}

\font\frtnsf = cmss12 scaled\magstep1
\font\twlsf = cmss10
\font\tensf = cmss9

\newfam\Scfam
\textfont\Scfam = \frtnsf
\scriptfont\Scfam = \twlsf
\scriptscriptfont\Scfam = \tensf 

\makeatletter \@addtoreset{equation}{section}
\def\theequation{\thesection.\arabic{equation}} \makeatother
\begin{document}

\sloppy \title { Cohomologies of the Poisson superalgebra on \\
(2,n)-superdimensional spaces. } \author { S.E.Konstein\thanks{E-mail:
konstein@lpi.ru}\ \ and I.V.Tyutin\thanks{E-mail: tyutin@lpi.ru} \thanks{
This work was supported by the RFBR (grants No.~02-01-00930 (I.T.) and
No.~02-02-17067 (S.K.)), by INTAS (grant No.~00-00-262 (I.T.)) and by the
grant LSS-1578.2003.2. } \\ { \sf \small I.E.Tamm Department of
Theoretical Physics,} \\ { \sf \small P. N. Lebedev Physical Institute,} \\
{ \sf \small 119991, Leninsky Prospect 53, Moscow, Russia.} } \date { }

\maketitle

\begin{abstract} { \footnotesize Cohomology spaces of the Poisson
superalgebra realized on smooth Grassmann-valued functions with compact
support on $\R^2$ are investigated under suitable continuity restrictions on
cochains. The zeroth, first and second cohomology spaces with coefficients
in the adjoint representation of the Poisson superalgebra are found for the
case of a constant nondegenerate Poisson superbracket.} \end{abstract}

\section{Introduction} \label{intr}

The hope to construct the quantum mechanics on nontrivial manifolds is
connected with geometrical or deformation quantization \cite{1} - \cite{4}.
The functions on the phase space are associated with the operators, and the
product and the commutator of the operators are described by associative
*-product and *-commutator of the functions. These *-product and *-commutator
are the deformations of usual product and of usual Poisson bracket.

To find the deformations of Poisson superalgebra, one should calculate the
second cohomology of the Poisson superalgebra.

In \cite{Zh}, the lower cohomologies (up to second) were calculated for the
Poisson algebra consisting of smooth complex-valued functions on ${\open
R}^{n}$. In \cite{Leites}, the deformations of the Lie superalgebra of
Hamiltonian vector fields with polynomial coefficients are investigated. The
pure Grassmannian case $n = 0$ is considered in \cite{Ty1} and \cite{Ty2}.

In \cite{n>2}, the lower cohomologies of the Poisson superalgebra of
Grassmann valued functions with compact support on ${\open R}^n$ in the
trivial (up to third cohomology) and adjoint representation (up to second
cohomology) for the case $n \ge 4$ are calculated. It occurred that the case
$n = 2$ needs a separate consideration which is provided in the present
paper. In \cite{central} we study the central extensions of the algebras
considered in \cite{n>2} and in this paper.

Let $\K$ be either $\R$ or $\oC$. We denote by $\EuScript D(\R^n)$ the space
of smooth $\K$-valued functions with compact support on $\R^n$. This space is
endowed by its standard topology: by definition, a sequence $\varphi_k\in
\EuScript D(\R^n)$ converges to $\varphi\in \EuScript D(\R^n)$ if
$\partial^\lambda\varphi_k$ converge uniformly to $\partial^\lambda\varphi$
for every multi-index $\lambda$, and the supports of all $\varphi_k$ are
contained in a fixed compact set. We set $$ \mathbf D^{n_-}_{n_ + } =
\EuScript D(\R^{n_ + })\otimes \G^{n_-},\quad \mathbf E^{n_-}_{n_ + } =
C^\infty(\R^{n_ + })\otimes \G^{n_-},\quad \mathbf D^{\prime n_-}_{n_ + } =
\EuScript D'(\R^{n_ + })\otimes \G^{n_-}, $$ where $\G^{n_-}$ is the
Grassmann algebra with $n_-$ generators and $\EuScript D'(\R^{n_ + })$ is the
space of continuous linear functionals on $\EuScript D(\R^{n_ + })$. The
generators of the Grassmann algebra (resp., the coordinates of the space
$\R^{n_ + }$) are denoted by $\xi^\alpha$, $\alpha = 1,\ldots,n_-$ (resp.,
$x^i$, $i = 1,\ldots, n_ + $). We shall also use collective supervariables
$z^A$ which are equal to $x^A$ for $A = 1,\ldots,n_ + $ and are equal to
$\xi^{A-n_ + }$ for $A = n_ + + 1,\ldots,n_ + + n_-$. The spaces $\mathbf
D^{n_-}_{n_ + }$, $\mathbf E^{n_-}_{n_ + }$, and $\mathbf D^{\prime n_-}_{n_
+ }$ possess a natural grading which is determined by that of the Grassmann
algebra. The parity of an element $f$ of these spaces is denoted by
$\varepsilon(f)$. We also set $\varepsilon_A = 0$ for $A = 1,\ldots, n_ + $
and $\varepsilon_A = 1$ for $A = n_ + + 1,\ldots, n_ + + n_-$.

Let $\partial/\partial z^A$ and $\overleftarrow{\partial}/\partial z^A$ be
the operators of the left and right differentiation. The Poisson bracket is
defined by the relation \begin{equation} \{f,g\}(z) =
f(z)\frac{\overleftarrow{\partial}}{\partial z^A} \omega^{AB}
\frac{\partial}{\partial z^B}g(z) = - \sigma(f,g)\{g,f\}(z),\label{3.0}
\end{equation} where $ \sigma(f,g) = (-1)^{\varepsilon(f)\varepsilon(g)}$ and
the symplectic metric $\omega^{AB} = (-1)^{\varepsilon_A \varepsilon_B}
\omega^{BA}$ is a constant invertible matrix. {}For definiteness, we choose
it in the form $$ \omega^{AB} = \left(\begin{array}{cc} \omega^{ij}&0 \\
0&\lambda_\alpha\delta^{\alpha\beta} \end{array} \right),\quad \lambda_\alpha
= \pm1,\ i,j = 1,...,n_ + ,\ \alpha,\beta = 1 + n_ + ,...,n_- + n_ +, $$
where $\omega^{ij}$ is the canonical symplectic form (if $\K = \oC$, then one
can choose $\lambda_\alpha = 1$). The nondegeneracy of the matrix
$\omega^{AB}$ implies, in particular, that $n_ + $ is even. The Poisson
superbracket satisfies the Jacobi identity \begin{equation}
\sigma(f,h)\{f,\{g,h\} \}(z) + \hbox{cycle}(f,g,h) = 0,\quad f,g,h\in
\mathbf E^{n_-}_{n_ + }. \label{3.0a} \end{equation} By Poisson
superalgebra, we mean the space $\mathbf D^{n_-}_{n_ + }$ with the Poisson
bracket~(\ref{3.0}) on it. The relations~(\ref{3.0}) and~(\ref{3.0a}) show
that this bracket indeed determines a Lie superalgebra structure on $\mathbf
D^{n_-}_{n_ + }$.

The integral on $\mathbf D^{n_-}_{n_ + }$ is defined by the relation $ \int
\di z\, f(z) = \int_{\R^{n_ + }} \di x\int \di\xi\, f(z), $ where the
integral on the Grassmann algebra is normed by the condition $\int \di\xi\,
\xi^1\ldots\xi^{n_-} = 1$. We identify $\G^{n_-}$ with its dual space
$\G^{\prime n_-}$ setting $f(g) = \int\di\xi\, f(\xi)g(\xi)$, $f,g\in
\G^{n_-}$. Correspondingly, $\mathbf D^{\prime n_-}_{n_ + }$, i.e., the
space of continuous linear functionals on $\mathbf D^{n_-}_{n_ + }$ is
identified with the space $\D^\prime(\R^{n_ + })\otimes \G^{n_-}$. As a rule,
the value $m(f)$ of a functional $m\in \mathbf D^{\prime n_-}_{n_ + }$ on a
test function $f\in \mathbf D^{n_-}_{n_ + }$ will be written in the
``integral'' form: $ m(f) = \int \di z\, m(z) f(z). $

Let $L$ be a Lie superalgebra acting in a $\Z_2$-graded space $V$ (the action
of $f\in L$ on $v\in V$ will be denoted by $f\cdot v$). The space $C_p(L, V)$
of $p$-cochains consists of all multilinear super antisymmetric mappings from
$L^p$ to $V$. The space $C_p(L, V)$ possesses a natural $\Z_2$-grading: by
definition, $M_p\in C_p(L, V)$ has the definite parity $\varepsilon(M_p)$ if
$$ \varepsilon(M_p(f_1,\ldots,f_p)) = \varepsilon(M_p) + \varepsilon(f_1) +
\ldots + \varepsilon(f_1) $$ for any $f_j\in L$ with parities
$\epsilon(f_j)$. The differential $\di_p^V$ is defined to be the linear
operator from $C_p(L, V)$ to $C_{p + 1}(L, V)$ such that \begin{align}
&&d_p^{V}M_p(f_1,...,f_{p + 1}) = - \sum_{j = 1}^{p + 1}(-1)^{j +
\varepsilon(f_j)|\varepsilon(f)|_{1,j-1} +
\varepsilon(f_j)\varepsilon_{M_p}}f_j\cdot M_p(f_{1},...,\breve{f}_j,...,f_{p
+ 1})- \nonumber \\ &&- \sum_{i<j}(-1)^{j +
\varepsilon(f_j)|\varepsilon(f)|_{i + 1,j-1}}
M_p(f_1,...f_{i-1},\{f_i,f_j\},f_{i + 1},...,\breve{f}_j,...,f_{p + 1}),
\label{diff} \end{align} for any $M_p\in C_p(L, V)$ and $f_1,\ldots f_{p + 1}
\in L$ having definite parities. Here the sign $\breve{}$ means that the
argument is omitted and the notation $$ |\varepsilon(f)|_{i,j} = \sum_{l =
i}^j\varepsilon(f_l) $$ has been used. The differential $\di^V$ is nilpotent
(see \cite{Schei97}), i.e., $\di^V_{p + 1} \di^V_p = 0$ for any $p =
0,1,\ldots$. The $p$-th cohomology space of the differential $\di_p^V$ will
be denoted by $H^p_V$.

We study the cohomologies of the Poisson algebra $\mathbf D^{n_-}_{n_ + }$ in
the following representations: \begin{enumerate} \item $V = \mathbf D^{\prime
n_-}_{n_ + }$ and $f\cdot m = \{f,m\}$ for any $f\in\mathbf D^{n_-}_{n_ + }$
and $m\in \mathbf D^{\prime n_-}_{n_ + }$. The space $C_p(\mathbf D^{n_-}_{n_
+ }, \mathbf D^{\prime n_-}_{n_ + })$ consists of separately continuous
antisymmetric multilinear mappings from $(\mathbf D^{n_-}_{n_ + })^p$ to
$\mathbf D^{\prime n_-}_{n_ + }$. The continuity of $M\in C_p(\mathbf
D^{n_-}_{n_ + }, \mathbf D^{\prime n_-}_{n_ + })$ means that the ($p +
1$)-form $$ (f_1,\ldots,f_{p + 1})\to \int \di z\, M(f_1,\ldots,f_p)(z)f_{p +
1}(z) $$ on $(\mathbf D^{n_-}_{n_ + })^{p + 1}$ is separately continuous. The
cohomology spaces will be denoted by $H^p_{\mathbf D'}$. \item $V = \mathbf
E^{n_-}_{n_ + }$ and $f\cdot m = \{f,m\}$ for every $f\in\mathbf D^{n_-}_{n_
+ }$ and $m\in \mathbf E^{n_-}_{n_ + }$. The space $C_p(\mathbf D^{n_-}_{n_ +
}, \mathbf E^{n_-}_{n_ + })$ is the subspace of $C_p(\mathbf D^{n_-}_{n_ + },
\mathbf D^{\prime n_-}_{n_ + })$ consisting of mappings taking values in
$\mathbf E^{n_-}_{n_ + }$. The cohomology spaces will be denoted by
$H^p_{\mathbf E}$. \item The adjoint representation: $V = \mathbf D^{n_-}_{n_
+ }$ and $f\cdot g = \{f,g\}$ for any $f,g\in\mathbf D^{n_-}_{n_ + }$. The
space $C_p(\mathbf D^{n_-}_{n_ + }, \mathbf D^{n_-}_{n_ + })$ is the subspace
of $C_p(\mathbf D^{n_-}_{n_ + }, \mathbf D^{\prime n_-}_{n_ + })$ consisting
of mappings taking values in $\mathbf D^{n_-}_{n_ + }$. The cohomology spaces
and the differentials will be denoted by $H^p_{\mathrm{ad}}$ and
$\di^{\mathrm{ad}}_p$ respectively.  \end{enumerate} {}For these
representations we shall denote the differentials by the same symbol
$\di^{\mathrm{ad}}_p$ as in the adjoint representation.  We shall call
p-cocycles $M_p^1,\ldots M_p^k$ independent if they give rise to linearly
independent elements in $H^p$. {}For a multilinear form $M_p$ taking values
in $\mathbf D^{n_-}_{n_ + }$, $\mathbf E^{n_-}_{n_ + }$, or $\mathbf
D^{\prime n_-}_{n_ + }$, we write $M_p(z|f_1,\ldots,f_p)$ instead of more
cumbersome $[M_p(f_1,\ldots,f_p)](z)$.

Below we assume that $n_ + = 2$.

The main results of this work are given by the following two theorems.

\theorem\label{th2} {\it \begin{enumerate} \item $H^0_{\mathbf D'} \simeq
H^0_{\mathbf E} \simeq \K$, the function $M_0(z)\equiv 1$ is a nontrivial
cocycle.

\item $H^1_{\mathbf D'} \simeq H^1_{\mathbf E} \simeq \K^2$, independent
nontrivial cocycles are given by $$ M^1_1(z|f) = \bar f,\quad M^2_1(z|f) =
\eE_z f(z), $$ where \footnote{The operator ${\cal E}_z$ is a derivation of
the Poisson superalgebra.} $\eE_z \stackrel {def} = 1-\frac 1 2 z^A \frac
{\partial} {\partial z^A} \,\,\,$ and $\bar f = \int du f(u)$.

\item\label{p23} Let the bilinear mappings $M_2^1$, $M_2^2$, $M_2^3$,
$M_2^4$, $N^E_2$, $L_2^0$, $L_2^1$, $L_2^2$, and $L_2^3$ from $(\mathbf
D^{n_-}_{2})^2$ to $\mathbf E^{n_-}_{2}$ be defined by the relations
\begin{eqnarray} M_2^1(z|f,g)& =
&f(z)\!\left(\frac{\overleftarrow{\partial}}{\partial z^A} \omega^{AB}
\frac{\partial}{\partial z^B} \right)^3\!g(z), \nonumber \\ M^2_2(z|f,g)& =
&\bar f \bar g, \nonumber \\ M^3_2(z|f,g)& = & \bar g \eE_z f(z)- \sigma(f,g)
\bar f \eE_z g(z), \nonumber \\ M^4_2(z|f,g)& = & \int du\left(f(u) {} \eE_u
g(u)- \sigma(f,g){g(u)}{} \eE_u f(u)\right), \nonumber \\ N^E_2(z|f,g)& = &
\Theta(z|\partial_{2}fg)-\Theta(z|f\partial_{2}g)-2(-1)^{n_-\varepsilon (f)}
\partial_{2}f(z)\Theta(z|g) + 2\Theta(z|f)\partial_{2}g(z) \nonumber \\
L^0_2(x|f,g)& = & N^E_2(x|f,g) + \frac{1}{2} \left ( x^{i} \partial_{i}f(x)
\right) g(x) -\frac{1}{2} f(x) \left ( x^i\partial_{i}g(x)\right),
\nonumber\\ L^1_2(z|f,g)& = & N^E_2(z|f,g)- \Delta (z|f) g(z) +
(-1)^{\varepsilon(f)} f(z)\Delta (z|g) \nonumber \\ && -\frac{2}{3}
(-1)^{\varepsilon(f)} \left ( \xi^1\partial_{\xi^1}f(z) \right) \Delta (z|g),
\nonumber\\ L^2_2(z|f,g)& = &N^E_2(z|f,g)-\Delta (z|f)g(z) + f(z)\Delta
(z|g) \nonumber\\ L^3_2(z|f,g)& = &N^E_2(z|f,g) -\Delta (z|f)g(z)+ 
(-1)^{\varepsilon (f)}f(z)\Delta (z|g) + \nonumber\\ && + \partial _{\xi
^{\alpha }}f(z)\Delta _{\alpha }(z|g)-(-1)^{\varepsilon (f)} \Delta _{\alpha
}(z|f)\partial _{\xi ^{\alpha }}g(z), \nonumber \end{eqnarray} where
\begin{eqnarray} \Theta(z|f) & \stackrel {def} = &\int du
\delta(x_1-u_1)\theta(x_2-u_2)f(u), \nonumber\\ \Delta (z|f) & \stackrel
{def} = &\int du \delta(x-y) f(u), \nonumber\\ \Delta _{\alpha}(z|f) &
\stackrel {def} = &\int du \eta_\alpha\delta(x-y) f(u), \end{eqnarray} and $z
= (x_1,x_2,\xi_1,\,...\,,\,\xi_{n_-})$, $u =
(y_1,y_2,\eta_1,\,...\,,\,\eta_{n_-})$.

If $n_-$ is even and $n_- \ne 0$, $n_- \ne 2$, $n_- \ne 6$ then $H^2_{\mathbf
D'} \simeq H^2_{\mathbf E} \simeq \K^2$ and the cochains $M^1_2$ and $M^3_2$
are independent nontrivial cocycles;

if $n_-$ is odd and $n_- \ne 1$, $n_- \ne 3$ then $H^2_{\mathbf D'} \simeq
H^2_{\mathbf E} \simeq \K^3$ and the cochains $M^1_2$, $M^2_2$ and $M^3_2$
are independent nontrivial cocycles;

if $n_- = 0$, then $H^2_{\mathbf D'} \simeq H^2_{\mathbf E} \simeq \K^3$ and
the cochains $M^1_2$, $M^3_2$ and $L^0_2$ are independent nontrivial
cocycles;

if $n_- = 1$, then $H^2_{\mathbf D'} \simeq H^2_{\mathbf E} \simeq \K^4$ and
the cochains $M^1_2$, $M^2_2$, $M^3_2$ and $L^1_2$ are independent nontrivial
cocycles;

if $n_- = 2$, then $H^2_{\mathbf D'} \simeq H^2_{\mathbf E} \simeq \K^3$ and
the cochains $M^1_2$, $M^3_2$ and $L^2_2$ are independent nontrivial
cocycles;

if $n_- = 3$, then $H^2_{\mathbf D'} \simeq H^2_{\mathbf E} \simeq \K^4$ and
the cochains $M^1_2$, $M^2_2$, $M^3_2$ and $L^3_2$ are independent nontrivial
cocycles;

if $ n_- = 6$, then $H^2_{\mathbf D'} \simeq H^2_{\mathbf E} \simeq \K^3$ and
the cochains $M^1_2$, $M^3_2$, and $M^4_2$ are independent nontrivial
cocycles. \end{enumerate} }

\theorem\label{th3}{\it \begin{enumerate} \item $H^0_{\mathrm{ad}} = 0$.

\item Let $V_1$ be the one-dimensional subspace of $C_1(\mathbf
D^{n_-}_{2},\mathbf D^{n_-}_{2})$ generated by the cocycle $M^2_1$ defined in
Theorem~\ref{th2}. Then there is a natural isomorphism $V_1\oplus (\mathbf
E^{n_-}_{2}/\mathbf D^{n_-}_{2}) \simeq H^1_{\mathrm{ad}}$ taking $(M_1,T)\in
V_1\oplus (\mathbf E^{n_-}_{2}/\mathbf D^{n_-}_{2})$ to the cohomology class
determined by the cocycle $M^2_1(z|f) + \{t(z),f(z)\}$, where $t\in \mathbf
E^{n_-}_{2}$ belongs to the equivalence class $T$.

\item Let bilinear mapping $N^E_2$ from $(\mathbf D^{n_-}_{2})^2$ to $\mathbf
E^{n_-}_{2}$ be defined in Theorem \ref{th2} and $$N_1(f) =
-2\Lambda(x_2)\int du \theta(x_1-y_1)f(u),$$ \noindent where $\Lambda\in
C^\infty({\mathbb R})$ be a function such that $\frac d {dx} \Lambda \in
\D(\R)$ and $\Lambda(-\infty) = 0,\ \Lambda( + \infty) = 1$.

Then bilinear mapping $N^D_2(f,g) = N^E_2(f,g) + d_1^{\mathrm{ad}}N_1(f,g)$
maps $(\mathbf D^{n_-}_{2})^2$ to $\mathbf D^{n_-}_{2}$.

\item Let the bilinear mappings $M_2^1$, $M_2^3$, $L_2^0$, $L_2^1$, $L_2^2$,
and $L_2^3$ from $(\mathbf D^{n_-}_{2})^2$ to $\mathbf D^{n_-}_{2}$ be
defined by Theorem~\ref{th2}.

Let $V^{n_-}_2$ be the subspace of $C_2(\mathbf D^{n_-}_{2},\mathbf
D^{n_-}_{2})$ generated by the cocycles $M^1_2$, $M^3_2$ and $L_2^{n_-} +
d_1^{\rm ad} N_1$ for $n_- = 0,1,2,3$, and by the cocycles $M^1_2$ and
$M^3_2$ for $n_- \ge 4$.

Then there is a natural isomorphism $V^{n_-}_2\oplus (\mathbf
E^{n_-}_{2}/\mathbf D^{n_-}_{2}) \simeq H^2_{\mathrm{ad}}$ taking $(M_2,T)\in
V^{n_-}_2\oplus (\mathbf E^{n_-}_{2}/\mathbf D^{n_-}_{2})$ to the cohomology
class determined by the cocycle $$ M_2(z|f,g)-\{t(z),f(z)\} \bar{g} +
\sigma(f,g)\{t(z),g(z)\} \bar{f}, $$ where $t\in \mathbf E^{n_-}_{2}$
belongs to the equivalence class $T$. \end{enumerate} }

\section{Preliminaries} \label{prel} We adopt the following multi-index
notation: $$ (z^A)^0\equiv1,\quad (z^A)^k\equiv z^{A_1} \cdots z^{A_k},\quad
k \geq 1, $$ $$ (\partial^z_A)^0\equiv1,\quad (\partial^z_A)^k\equiv
\frac{\partial}{\partial z^{A_1}} \cdots \frac{\partial}{\partial
z^{A_k}},\quad k \geq 1, $$ $$
(\overleftarrow{\partial}^z_A)^0\equiv1,\quad(\overleftarrow{\partial}^z_A)^k
\equiv \frac{\overleftarrow{\partial}}{\partial z^{A_k}} \cdots
\frac{\overleftarrow{\partial}}{\partial z^{A_1}},\quad k \geq 1, $$ Besides,
we introduce the following notation \begin{eqnarray} \DD_z^A \stackrel {def}
= \frac {\overleftarrow {\partial}}{\partial z^B} \omega^{BA}. \end{eqnarray}
{}Further, $$ T^{\ldots(A)_k\ldots } \equiv T^{\ldots A_1\ldots A_k\ldots
},\quad T^{\ldots A_iA_{i + 1} \ldots } = \sigma(A_i,A_{i + 1})T^{\ldots A_{i
+ 1}A_i\ldots },\quad i = 1,\ldots,k-1. $$

The $\delta$-function is normed by the relation $ \int dz^{\prime } \delta
(z^{\prime }-z)f(z^{\prime }) = \int f(z^{\prime })\delta (z-z^{\prime
})dz^{\prime } = f(z). $

We denote by $M_p(\ldots)$ the separately continuous $p$-linear forms on
$(\mathbf D^{n_-}_{2})^p$. Thus, the arguments of these functionals are the
functions $f(z)$ of the form \begin{equation} \label{dec} f(z) = \sum_{k =
0}^{n_-} f_{(\alpha)_k}(x)(\xi^\alpha)^k,\quad f_{(\alpha)_k}(x)\in
\D(\R^{2}). \end{equation} It can be easily proved that such multilinear
forms can be written in the integral form \begin{equation}
M_p(z|f_1,\ldots,f_p) = \int dz_p\cdots dz_1m_p(z|z_1,\ldots,z_p)
f_1(z_1)\cdots f_p(z_p),\;p = 1,2,... \end{equation} It follows from the
antisymmetry properties of the forms $M_p$ that the corresponding kernels
$m_p$ have the following properties: \begin{eqnarray}
\varepsilon(m_p(z|z_1,\ldots,z_p)) = pn_{-} + \varepsilon_{M_p}, \nonumber \\
m_p(|z_1\ldots z_i,z_{i + 1} \ldots z_p) = -(-1)^{n_-} m_p(z|z_{1} \ldots
z_{i + 1},z_i\ldots z_p). \label{3.2a} \end{eqnarray}

The support $ \supp(f) \subset \R^{2}$ of a test function $f\in \mathbf
D^{n_-}_{2}$ is defined to be the union of the supports of all
$f_{(\alpha)_k}$ entering in the decomposition~(\ref{dec}). {}For $m\in
\mathbf D^{\prime n_-}_{2}$, the set $ \supp(m)$ is defined analogously.

The sign $\hat{}$ (hat) over the form $M$ or over its kernel $m(z|u,v,...)$
means that the kernel $m$ is considered off the diagonals $z = u$, $z = v$,
...

The following low degree filtrations $\oP_{p_1,p_2,...,p_s}$ of the
polynomials we define as: \definition \begin{eqnarray}
&&\oP_{p_1,p_2,...,p_s} = \{f(k_1,k_2,...,k_s)\in \K [k_1,...,k_s]:
\nonumber\\ && \ \ \ \ \ \ \exists g \in \K [\alpha_1,\alpha_2,...,\alpha_s,
k_1,...,k_s]\ \ f(\alpha_1 k_1,\alpha_2 k_2,...) = \alpha_1^{p_1}
\alpha_2^{p_2}...\alpha_s^{p_s} g\}, \end{eqnarray} where $k_1$, $k_2$, ... ,
$k_s$ are some sets of supervariables, and $\alpha_1$, ... , $\alpha_s$ are
some even variables.

Evidently, $\oP_{p_1,p_2...} \subset \oP_{q_1,q_2...}$ if $p_i \geq q_i$.  It
is clear also, that if $f\in \oP_{p_1,p_2,...}$ and $g\in \oP_{q_1,q_2,...}$
then $fg\in \oP_{p_1 + q_1,p_2 + q_2,...}$.

\section{Cohomologies in the adjoint representation} \label{adjo}

In this and following section, we prove Theorems~\ref{th2} and~\ref{th3}. We
assume that the forms under consideration take values in the space $\cal A$,
where $\cal A$ is one of the spaces $\mathbf D^{\prime n_-}_{2}$, $\mathbf
E^{n_-}_{2}$, or $\mathbf D^{n_-}_{2}$.

\subsection{$H^0_{\rm ad}$} \label{Had0}

The proof of Theorems~\ref{th2} and~\ref{th3} for describing the zeroth
cohomologies in the adjoint representation is the same as for the case $n_ +
\geq 4$ and is presented in \cite{n>2}.

\subsection{$H^1_{\rm ad}$} \label{Had1}

The proof of Theorems~\ref{th2} and~\ref{th3} for describing the first
cohomologies in the adjoint representation is the same as for the case $n_ +
\geq 4$ and is presented in \cite{n>2}.

\subsection{$H^2_{\rm ad}$} \label{Had2}

{}For the bilinear form $$ M_2(z|f,g) = \int dvdum_2(z|u,v)f(u)g(v)\in{\cal
A}, $$ the cohomology equation has the form \begin{eqnarray} 0& =
&d_{2}^{\mathrm{ad}}M_{2}(z|f,g,h) = \nonumber \\ & = &
\sigma(g,h)\{M_{2}(z|f,h),g(z)\}-\{M_{2}(z|f,g),h(z)\} - \sigma (f,gh)
\{M_{2}(z|g,h),f(z)\}- \nonumber \\ &&-M_{2}(z|\{f,g\},h) + \sigma
(g,h)M_{2}(z|\{f,h\},g) + M_{2}(z|f,\{g,h\}). \label{5.4} \end{eqnarray}

The solution of (\ref{5.4}) can be presented as a sequence of the following
propositions:

\proposition\label{q4.2.1} {\it $M_2(z|f,g) = M_{2|1}(z|f,g) +
M_{2|2}^1(z|f,g) + M_{2|2}^2(z|f,g)$, where \begin{eqnarray} M_{2|1}(z|f,g) =
c\bar{f} \bar{g},\; \varepsilon(c) = \varepsilon(M_2),\quad d_{2}^{\rm
ad}M_{2|1}(z|f,g,h) = 0, \label{5.4b} \\ M_{2|2}^{1}(z|f,g) = \sum_{q = 0}^Q
m^{1(A)_q}(z|[(\partial^z_A)^qf(z)]g- \sigma(f,g)[(\partial^z_A)^qg(z)]f),
\label{5.4d} \\ M_{2|2}^{2}(z|f,g) = \sum_{q =
0}^Qm^{2(A)_q}(z|[(\partial_A)^qf]g- \sigma(f,g) (\partial_A)^qg]f),
\label{5.4h} \end{eqnarray} and the parities of the linear forms
$m^{1,2(A)_q}(z|\cdot)$ are equal to \begin{eqnarray} \varepsilon
(m^{1,2(A)_q}) = \varepsilon(M_2) + |\varepsilon_A|_{1,q} + n_-.
\end{eqnarray} Due to antisymmetry of bilinear form the constant $c$ can be
nonzero only if $n_-$ is odd.}

Let us denote the space of all bilinear forms with the representation
(\ref{5.4d}) as ${\cal M}_1$ and with the representation (\ref{5.4h}) as
${\cal M}_2$.

The space ${\cal M}_0 = {\cal M}_1 \bigcap {\cal M}_1 \ne \{0\}$ is called in
this paper the space of local bilinear forms. So the decomposition of this
Proposition is unique up to local bilinear forms.

To prove this Proposition it suffices to consider Eq.~(\ref{5.4}) in the
following two domains: \begin{eqnarray} && z\bigcap\left[{\rm
supp}(f)\bigcup{\rm supp}(g)\bigcup{\rm supp}(h)\right] = {\rm
supp}(f)\bigcap\left[{\rm supp}(g)\bigcup{\rm supp}(h)\right] = \varnothing
\mbox { and } \nonumber\\ && z\bigcap \! \left[{\rm supp}(f)\bigcup{\rm
supp}(g)\right]\!\! = \! {\rm supp}(f)\bigcap \! \left[{\rm
supp}(g)\bigcup{\rm supp}(h)\right]\!\! = \! {\rm supp}(g)\bigcap{\rm
supp}(h) \! = \! \varnothing. \nonumber \end{eqnarray}

\proposition\label{q4.2.2} {\it \begin{eqnarray} M_{2|2}^2(z|f,g)& =
&\delta_{n_-,6} \mu(z)\int du\left( u^A\frac{\partial f(u)}{\partial
u^A}g(u)- \sigma(f,g)u^A \frac{\partial g(u)}{\partial u^A}f(u)\right)
\nonumber \\ & + & V_2(z)\Theta(z|\partial_2f\,g-f\,\partial_2 g) +
d_{1}^{{\rm ad}} \zeta(z|f,g) + M_{2\,\mbox{loc}} \nonumber \end{eqnarray} }
{\it Proof.} Consider Eq.~(\ref{5.4}) into the domain $ z\bigcap\left[{\rm
supp}(f)\bigcup{\rm supp}(g)\bigcup{\rm supp}(h)\right] = \varnothing $,
where $M_{2|2}^1(z|f,g)$ is equal to zero identically. Using the results of
the subsection~3.2 of \cite{n>2} we obtain $$ \hat{M}_{2|2}^2(z|f,g) = \int
du\hat{m}^{2A}(z|u)\left( \frac{\partial f(u)}{\partial u^A}g(u)-
\sigma(f,g)\frac{\partial g(u)}{\partial u^A}f(u)\right) $$ and the vector
$\hat{m}^{2A}(z|u)$ satisfies the equation \be \label{4.2.2}
2\hat{m}^{2A}(z|u)\frac{\overleftarrow{\partial}}{\partial u^C} \omega^{CB}-
2 \sigma(A,B)\hat{m}^{2B}(z|u)\frac{\overleftarrow{\partial}}{\partial u^C}
\omega^{CA} + \mu(z)\omega ^{AB} = 0, \ee where $ \mu(z) = m^{2A}(z|u)
\frac{\overleftarrow{\partial}}{\partial u^A}(-1)^{\varepsilon_A},\;
\varepsilon(\mu) = \varepsilon(M_2) + n_-, $ does not depend on $u$. It
follows from this equation that $ (6-n_{-})\mu(z) = 0$.

So, either $n_- = 6$ or $\mu(z) = 0$.

The partial solution of (\ref{4.2.2}) has the simple form
$\hat{m}^{2A}_{partial}(z|u) = -\frac 1 4 \mu (z) u^A$.

Let us find the solution of homogeneous equation \be \label{4.2.2.0}
\hat{m}^{2A}(z|u)\frac{\overleftarrow{\partial}}{\partial u^C} \omega^{CB}-
\sigma(A,B)\hat{m}^{2B}(z|u)\frac{\overleftarrow{\partial}}{\partial u^C}
\omega^{CA} = 0, \ee

Consider 3 cases separately: {\bf i)} $A = \alpha>2$, $B = \beta>2$, {\bf
ii)} $A = i \le 2$, $B = \beta > 2$, and {\bf iii)} $A = i\le 2$, $B = j\le
2$.

{\bf i)}

Let $A = \alpha$, $B = \beta$, $$
\hat{m}^{2\alpha}(z|u)\overleftarrow{D}_u^{\beta} + \hat
{m}^{2\beta}(z|u)\overleftarrow{D}_u^{\alpha} = 0. $$ Introduce the tensor $$
T^{\alpha\beta}(z|u) = m^{2\alpha}(z|u)\overleftarrow{D}_u^{\beta} +
m^{2\beta}(z|u)\overleftarrow{D}_u^{\alpha}. $$

Since $\hat T^{\alpha\beta}(z|u) = 0$ we have $$ T^{\alpha\beta}(z|u) =
\sum_{q}t_{\alpha\beta}^{(l)_{q}}(z|\eta)(\partial_{l} )^{q} \delta(x-y). $$

The identity $T^{\alpha\beta} \overleftarrow{D}_u^{ \gamma} +
cycle(\alpha,\beta, \gamma) = 0$ gives
$t_{\alpha\beta}^{(l)_{q}}(z|\eta)\overleftarrow{D}_u^{ \gamma} +
\mathrm{cycle}(\alpha,\beta, \gamma) = 0$ and as a consequence $
t_{\alpha\beta}^{(l)_{q}}(z|\eta) =
t_{\alpha}^{(l)_{q}}(z|\eta)\overleftarrow {D}_u^{\beta} +
t_{\beta}^{(l)_{q}}(z|\eta)\overleftarrow {D}_u^{\alpha}, $ which implies
\be\label{albe} m^{2\alpha}(z|u) = m^{\prime2}(z|u)\overleftarrow{D}_u^\alpha
+ \sum_{q}t_{\alpha}^{(l)_{q}}(z|\eta)(\partial_{l})^{q} \delta(x-y) \ee
with some vectors $m^{\prime2}(z|u)$ and $t_{\alpha}^{(l)_{q}}(z|\eta)$.

{\bf ii)} \nopagebreak

Let $A = i$, $B = \beta$. Then, using (\ref{albe}), we can write $$
\lbrack\hat{m}^{2i}(z|u)-\hat{m}^{\prime2}(z|u)\overleftarrow{D}_u^{i}]
\overleftarrow{D}_u^{\beta} = 0. $$ Introduce the tensor $ T_{i\beta}(z|u) =
[m^{2i}(z|u)-m^{\prime2}(z|u)\overleftarrow{\partial} _{j}
\omega^{ji}]\overleftarrow{\partial}_{\beta} \lambda_{\beta} = \sum
_{q}t_{i\beta}^{(l)_{q}}(z|\eta)(\partial_{l})^{q} \delta(x-y). $ The
identity $ t_{i\beta}^{(l)_{q}}(z|\eta)\overleftarrow{D}_u^{\alpha} +
t_{i\alpha}^{(l)_{q}}(z|\eta)\overleftarrow{D}_u^{\beta} = 0$ gives
$t_{i\beta}^{(l)_{q}}(z|\eta) = t_{i}^{(l)_{q}}(z|\eta)
\overleftarrow{D}_u^{\beta}$, which implies $$ \lbrack
m^{2i}(z|u)-m^{2}(z|u)\overleftarrow{D}_u^{i}-
\sum_{q}t_{i}^{(l)_{q}}(z|\eta)(\partial_{l})^{q} \delta(x-y)]\overleftarrow
{D}_u^{\beta} = 0, $$ and, in its turn $$ m^{2i}(z|u) =
m^{\prime2}(z|u)\overleftarrow{D}_u^{i} +
\sum_{q}t_{i}^{(l)_{q}}(z|\eta)(\partial_{l})^{q} \delta(x-y) +
\sigma^{i}(z|y) $$ with some vectors $t_{i}^{(l)_{q}}(z|\eta)$ and $
\sigma^{i}(z|y)$.

So \be\label{m2a} m^{2A}(z|u) = m^{\prime2}(z|u)\overleftarrow{D}_u^{A} +
\delta_{Ai} \sigma^{i}(z|y) +
\sum_{q}t_{A}^{(l)_{q}}(z|\eta)(\partial_{l})^{q} \delta(x-y). \ee

{\bf iii)}

Let $A = i$, $B = j$. It follows from (\ref{m2a}) that $ \hat{
\sigma}^{i}(z|y)\overleftarrow{D}_y^{j}- \hat{
\sigma}^{j}(z|y)\overleftarrow{D}_y^{i} = 0$ and $\hat{ \sigma}^{i}(z|y)\frac
{\overleftarrow{\partial}}{\partial y^i} = 0. $ Introduce vector $$
T^{i}(z|y) = \sigma^{i}(z|y)-t(z|y)\overleftarrow{D}_y^{i}, \;t(z|y) =
\sigma^{1}(z|y)\frac{1}{\overleftarrow{D}_y^1}. $$ Evidently, $ T^{1}(z|y) =
0$ and $\hat{T}^{i}(z|y)\overleftarrow{\partial}_{y^i} = \hat{T}
^{2}(z|y)\overleftarrow{\partial}_{y^2} = 0$. So $$ \lbrack
\sigma^{2}(z|y)-t(z|y)\overleftarrow{\partial}_{y^1}]\overleftarrow
{\partial}_{y^2} = A(z|y) = \sum_{p,q \geq 0}A^{pq}(z)\partial_{1}^{p}
\partial_{2} ^{q} \delta(x-y). $$ Represent $A(z|y)$ in the form
\begin{align*} A(z|y) & = [A_{1}(z|y) +
A_{2}(z|y)\overleftarrow{\partial}_{y^1}-V_{2}
(z)\delta(x_{1}-y_{1})\theta(x_{2}-y_{2})]\overleftarrow{\partial}_{y^2},\\
A_{1}(z|y) & = \sum_{p,q \geq 0}A^{\prime pq}(z)\partial_{y^1}^{p}
\partial_{y^2} ^{q} \delta(x-y) = \mathrm{loc},\\ A_{2}(z|y) & = \sum_{p \geq
0}A^{p}(z)\partial_{1}^{p} \partial_{2}^{q}
\delta(x_{1}-y_{1})\theta(x_{2}-y_{2}),\;A_{2}(z|y)\overleftarrow{\partial
}_{y^2} = \sum_{p \geq 0}A^{\prime p}(z)\partial_{1}^{p} \partial_{2}^{q}
\delta(x-y) = \mathrm{loc}. \end{align*} Then $ \sigma^2$ acquires the form
$$
\sigma^{2}(z|y)-t(z|y)\overleftarrow{\partial}_{1}-A_{2}(z|y)\overleftarrow
{\partial}_{1}-A_{3}(z)\delta(x_{1}-y_{1})\theta(x_{2}-y_{2})-A_{1} (z|y) =
b(z|y_{1}) = a(z|y_{1})\overleftarrow{\partial}_{1}. $$ Thus, we obtain, that
$ \sigma^{i}(z|y)$ can be represented in the form \begin{align*}
\sigma^{i}(z|y) & = \sigma(z|y)\overleftarrow{D}_u^{i} + \left(
\begin{array} [c]{c} 0\\ V_{2}(z)\delta(x_{1}-y_{1})\theta(x_{2}-y_{2})
\end{array} \right) + \sigma^i_{\mathrm{loc}},\\ \sigma(z|y) & = t(z|y) +
A_{2}(z|y) + a(z|y_{1}).  \end{align*}

{}Finally, \begin{align*} m^{2A}(z|u) & = m^{2}(z|u)\overleftarrow{D}_u^{A} +
\delta _{A2}V_{2}(z)\delta(x_{1}-y_{1})\theta(x_{2}-y_{2}) +
\sum_{q}t_{A}^{(l)_{q} }(z|\eta)(\partial_{l})^{q} \delta(x-y), \end{align*}
which gives the statement of the Proposition.

Analogously one can prove the following

\proposition\label{q4.2.3} {\it The linear forms $m^{1(A)_q}(z|\cdot )$ from
Proposition \ref{q4.2.1} have the kernels of the form

\begin{eqnarray} \hat{m}^{1(A)_0}(z|u) = a(u);\ \;\hat{m}^{1(A)_q}(z|u) = 0,
\mbox{ for }q \geq 2, \nonumber\\ m^{1A}(z|u) = -\frac{1}{2}a(u)z^{A} +
m^1(z|u) {\overleftarrow{D}}_z^A \nonumber \\ + \delta
_{A2}V_{1}(u)\delta(x_{1}-y_{1})\theta(x_{2}-y_{2}) + m^1_{\rm loc},
\nonumber \\ \varepsilon(m^1) = \varepsilon(M_2) + n_-, \nonumber
\end{eqnarray} where $a(u)$, $V_{1}(u)$, $m^1(z|u)$ are some distributions
and $m^1_{\rm loc}$ is a distribution with support in the plane $z = u$.}

Thus we obtained that the cocycle $M_2(z|f,g)$ has the form \begin{align*}
M_{2}(z|f,g) & = M_{2|2}(z|f,g) + M_{2|3}(z|f,g) + M_{2|4}(z|f,g) + M_{2|5}
(z|f,g) + \\ & + M_{2|8}(z|f,g) + M_{2|\mathrm{loc}}(z|f,g) +
d_{1}^{\mathrm{ad}} \zeta ^{2}(z|f,g),\\ \end{align*} where \begin{align*}
M_{2|2}(z|f,g) & = c_{2} \bar{f} \bar{g}, \\ M_{2|3}(z|f,g)& = \int
dua(u)\left[ f(u) \eE_z g(z)- \sigma(f,g) g(u)\eE_z f(z)\right],\\
M_{2|4}(z|f,g) & = \delta_{n_{-},\,6} \mu(z)\int du[\{u^{A} \partial
_{A}f(u)\}g(u)-f(u)u^{A} \partial_{A}g(u)],\\ M_{2|5}(z|f,g) & =
V_{2}(z)[\Theta(z|\partial_{2}fg) \! - \! \Theta(z|f\partial _{2}g)] +
(-1)^{\varepsilon(f)\varepsilon_{M_{2}}} \partial_{2}f(z)\Theta (x|V_{1}g) \!
- \! \Theta(z|V_{1}f)\partial_{2}g(z),\\ M_{2|8}(z|f,g) & = \sigma(f,g)
g(u)\{m^{1} (z|g),f(z)\}-\{m^{1}(z|f),g(z)\}. \end{align*}

\proposition\label{q4.2.4} {\it The parameters $\mu(z)$ and $V_2$ from
Proposition \ref{q4.2.2}, and the parameters $a(u)$, $V_1$, and $ m^1(z,u)$
from Proposition \ref{q4.2.3} have the following form: $a(u) = c_3 = {\rm
const}$, $\mu(z) = c_4 = {\rm const}$, $V_2 = c_5$, $V_1 = -2c_5$, $m^1(z|u)
= \sum_{p,q \geq 0}B^{pq}(\xi,u)(\partial_1)^p(\partial^q_2)^q \delta(x-y) +
b(z) + c(u)$ }

{\it Proof.} The cohomological equation (\ref{5.4}) reduces now to
\begin{align} & \int du a(u)\{f(u),g(u)\} \eE_z h(z) + \nonumber\\ & +
\sigma(M_2(f,g),h) \delta_{n_{ + } + 4,n_{-}} \{h(z),\mu(z)\} \int
du[\{u^{A} \partial_{A}f(u)\}g(u)-f(u)u^{A} \partial_{A}g(u)] + \nonumber \\
& + \{\hat{m}^{1}(z|\{f,g\}),h(z)\} + \Theta(z|V_{1}
\{f,g\})\partial_{2}h(z) + \nonumber \\ & \, + \sigma(M_2(f,g),h)
\{h(z),V_{2}(z)\}[\Theta(z|\partial _{2}f g) -\Theta(z|f\partial_{2}g)] =
0.\label{24} \end{align} in the domain $ \left[z\bigcup{\rm
supp}(h)\right]\bigcap\left[{\rm supp}(f)\bigcup {\rm supp}(g)\right] =
\varnothing$.

Choosing $h(u) = {\rm const}$ in some neighborhood of the point $z$ we obtain
$a(u) = {\rm const}$. Choosing $f(u) = {\rm const}$ in some neighborhood of
${\rm supp}(g)$ we obtain $\mu(z) = {\rm const}$. As a consequence, we obtain
$$ \partial_{2}h(z)\Theta(z|V_{1} \{f,g\}) +
\{h(z),V_{2}(z)\}[\Theta(z|\partial
_{2}fg)-\Theta(z|f\partial_{2}g)]-\{h(z),\hat{m}^{1}(z|\{f,g\})\} = 0, $$
which implies in its turn \begin{align} &
\delta_{A,2}[\delta(x_{1}-y_{1})\theta(x_{2}-y_{2})V_{1}(u)]\overleftarrow
{\partial}_{D} \omega^{DB} + 2\omega^{AC} \partial_{C}[V_{2}(z)\delta(x_{1}
-y_{1})\theta(x_{2}-y_{2})]\delta_{B,2}- \nonumber\\ & \,-\omega^{AC}
\partial_{C} \hat{m}^{1}(z|u)\overleftarrow{\partial}_{D} \omega^{DB} =
0.\label{or} \end{align} Considering all the combinations:  $A = 1,2,\alpha$,
$B = 1,2,\beta$ (see Appendix \ref{combi}), we obtain the result of the
Proposition.

When one substitutes the expression for $m^1(z|u)$ into the expression for
$M^1_{2|2}$, one find that $c(u)$ is cancelled and that $b(z)$ gives rise to
pure differential only.

The results of Propositions \ref{q4.2.1} - \ref{q4.2.4} can be summarized in
the following

\lemma\label{l4.2.1} {\it The solution of the equation (\ref{5.4}) has the
following form \begin{align*} & M_{2}(z|f,g) = c_{2}m_{2|2}(z|f,g) +
c_{3}m_{2|3}(z|f,g) + \delta_{n_{-},\,6} c_{4}m_{2|4}(z|f,g) + \\ & +
c_{5}m_{2|5}^{\prime}(z|f,g) + M_{2|\mathrm{loc}}(z|f,g) +
d_{1}^{\mathrm{ad}} \zeta^{2}(z|f,g), \end{align*} where \begin{align*}
m_{2|2}(z|f,g) & = \bar{f} \bar{g},\\ m_{2|3}(z|f,g) & = \bar{f} \eE_z
g(z)-(-1)^{\varepsilon(f)\varepsilon(g)} \bar{g} \eE_z f(z),\\ m_{2|4}(z|f,g)
& = 2\int du[f(u)\eE_u g(u) -\eE_u f(u) g(u)],\\ m_{2|5}^{\prime}(z|f,g) & =
\Theta(z|\partial_{2}fg)-\Theta(z|f\partial _{2}g)-2(-1)^{n_{-}
\varepsilon(f)} \partial_{2}f(z)\Theta(z|g) + 2\Theta (x|f)\partial_{2}g(z),
\end{align*} $c_i$ are some constants, and $c_2 = 0$ if $n_-$ is even.}

The forms $m_{2|2}(z|f,g)$, $m_{2|3}(z|f,g)$, and $\delta_{n_{-},\,6}
m_{2|4}(z|f,g)$ satisfy the cohomological equation.

\section{Local cohomological equation }

Since the bilinear forms $m_{2|2}$, $m_{2|3}$, and $\delta_{n_{-},\,6}
m_{2|4}$ are cocycles, the cohomology equation is reduced to \begin{equation}
\label{5.6} d_{2}^{{\rm ad}}M_{2|{\rm loc}}(z|f,g,h) = -c_5 d_{2}^{{\rm
ad}}m_{2|5}^\prime(z|f,g,h). \end{equation}

The local bilinear form has the following form \begin{eqnarray}
\label{locform} M_{2|{\rm loc}}(z|f,g) = \sum_{p,q =
0}^N(-1)^{\varepsilon(f)|\varepsilon_B|_{1,q}}m^{(A)_p|(B)_q}(z)
[(\partial^z_A)^pf(z)](\partial^z_B)^qg(z), \\ m^{(B)_q|(A)_p} = -
(-1)^{|\varepsilon_A|_{1,p}|\varepsilon_B|_{1,q}}m^{(A)_p|(B)_q},\ \ \
m^{(A)_0|(B)_0} = 0,\\ \varepsilon(m^{(A)_p|(B)_q}) = \varepsilon(M_2) +
|\varepsilon_A|_{1,p} + |\varepsilon_B|_{1,q}. \nonumber \end{eqnarray}

\subsection{Equation for local form}

One can observe that $d_{2}^{\mathrm{ad}}m_{2|5}^\prime(z|f,g,h)$ is a local
form. So we can use the method of \cite{Zh} to solve Eq.~(\ref{5.6}). Let
$x\in U$, where $U\in {\fam\openfam R}^{2}$ is some bounded domain. Consider
Eq.~(\ref{5.6}) for the functions of the form \begin{equation*} f(z) =
e^{z^Ap_A}f_0(z),\quad g(z) = e^{z^Aq_A}g_0(z),\quad h(z) =
e^{z^Ar_A}h_0(z),\quad\varepsilon(k_A) = \varepsilon(l_A) = \varepsilon(r_A)
= \varepsilon_A, \end{equation*} where $p = (a,\pi)$, $q = (b, \gamma)$ and
$r = (c,\rho)$ are new supervariables, $ f_0(z) = g_0(z) = h_0(z) = 1$ if
$x\in U$ ($z = (x,\xi)$). So, if $x\in U$ then Eq.~(\ref{5.6}) takes the form

\begin{align} \label{loceq} &[p,q]\Phi (z,p,q,r) + \{{}F(z,q,r), zp\} +
\mbox{cycle}(p,q,r) = \nonumber\\ &\ \ = \sigma c_{5}([p,q]\{\delta (\pi +
\rho ) + \delta ( \gamma + \rho )-2\delta (\rho )-\delta (\pi + \gamma +
\rho )\} + \nonumber \\ &\ \ \ + [\pi, \gamma] \{\delta (\pi + \rho ) +
\delta ( \gamma + \rho )-2\delta (\rho )- \frac{2}{3} \delta (\pi + \gamma +
\rho )\} + \mbox{cycle}(p,q,r), \end{align} where $ {}F(z,p,q) = \sum_{k,l =
0}^{N}m^{(A)_{k}|(B)_{l}}(z)(p_{A})^{k}(q_{B})^{l} = -{}F(z,q,p)$,
\begin{align*} \Phi(z,p,q,r) & = {}F(z,p + q,r)-{}F(z,p,r)-{}F(z,q,r) =
\Phi(z,q,p,r), \\ [a,b] & \stackrel {def} = a_{i} \omega^{ij}b_{j},\;[\pi,
\gamma] \stackrel {def} = \pi_{\alpha } \lambda_{\alpha}
\gamma_{\alpha},\;[p,q] \overset{def}{ = }(-1)^{\varepsilon _{A}} \omega
^{AB}p_{A}q_{B} = [a,b]-[\pi, \gamma], \end{align*} and $ \sigma =
(-1)^{n_-(n_- -1)/2}$.

Let \begin{eqnarray} \Psi (z|p,q,r) = \psi (z|p + q,r)-\psi (z|p,r)-\psi
(z|q,r), \label{generala} \\ \psi (z|p,q) = \sum_{k,l =
0}^{N}m(z)^{(A)_{k}|(B)_{l}}(p_{A})^{k}(q_{B})^{l} = -\psi (z|q,p).
\label{generalb} \end{eqnarray} be the solution of the homogeneous part of
Eq.~(\ref{loceq}): \begin{equation} \label{general_0} [p,q]\Psi (z|p,q,r) +
\{\psi (z|p,q),zr\} + \mathrm{cycle}(p,q,r) = 0. \end{equation}

This homogeneous equation is solved in Appendix \ref{hom}, and, up to
coboundary, its solution has the form \begin{equation} \label{5.13} \psi
(z|p,q) = c [p,q]^3, \end{equation} where $c$ is an arbitrary constant.

Now let us find the partial solutions of Eq.~(\ref{loceq}) for different
values of $n_-$.

\subsection{Partial solution of Equation (\protect\ref{loceq}) for the case
$\mathbf {n_- = 0.}$} Eq.~(\ref{loceq}) reduces to \begin{align*} &
[a,b]\Phi(x,a,b,c) + [b,c]\Phi(x,b,c,a) + [c,a]\Phi(x,c,a,b) + \\ \, & \, +
\{{}F(x,b,c),xa\} + \{{}F(x,c,a),xb\} + \{{}F(x,a,b),xc\} = \\ & \, =
-c_{5}([a,b] + [b,c] + [c,a]) \end{align*} and has a partial solution
\begin{equation} \label{0} {}F(x,a,b) = \frac{c_{5}}{2}(xa-xb).
\end{equation}

\subsection{Partial solution of Equation (\protect\ref{loceq}) for the case
$\mathbf {n_- = 1.}$} Eq.~(\ref{loceq}) reduces to \begin{align*} &
[p,q]\Phi(z,p,q,r) + [q,r]\Phi(z,q,r,p) + [r,p]\Phi(z,r,p,q) + \\ \, & \, +
\{{}F(z,q,r),zp\} + \{{}F(z,r,p),zq\} + \{{}F(z,p,q),zr\} = \\ & \, =
-c_{5}([p,q]\rho_1 + [q,r]\pi_1 + [r,p] \gamma_1 + 2\lambda_1\pi_1
\gamma_1\rho_1 ) \end{align*} and has a partial solution $$ {}F(z,p,q) =
c_{5} \Big( \gamma_1-\pi_1 -\frac {2} 3 \xi^1\pi_1 \gamma_1\Big), $$

\subsection{Equation (\protect\ref{loceq}) for $\mathbf{ n_- \ge 2}$}

Let ${}F(z,p,q) = c_{5} \sigma \lbrack \delta ( \gamma )-\delta (\pi )] +
{}F_{1}(z,p,q)$.

The cohomological equation (\ref{loceq}) for shifted form takes the form

\begin{align} & [p,q]\Phi_1 (z,p,q,r) + \{{}F_1(z,q,r),zp\} +
\mathrm{cycle}(p,q,r) = \notag \\ & \, = \sigma c_{5} \left( \aleph ([p,q] +
[q,r] + [r,p]) + 2\aleph ^{\alpha } \pi _{\alpha } \gamma _{\alpha } \rho
_{\alpha } \lambda_\alpha \right), \label{CE3} \end{align} where
\begin{equation} \aleph (\pi , \gamma ,\rho ) = \delta (\pi + \rho ) + \delta
( \gamma + \rho ) + \delta ( \gamma + \pi )-\delta (\pi )-\delta ( \gamma
)-\delta (\rho )-\delta (\pi + \gamma + \rho ), \label{alef} \end{equation}
\begin{equation*} \aleph ^{\alpha }(\pi , \gamma ,\rho ) = \delta _{\alpha
}(\pi + \rho ) + \delta _{\alpha }( \gamma + \rho ) + \delta _{\alpha }(
\gamma + \pi )-\delta _{\alpha }(\pi )-\delta _{\alpha }( \gamma )-\delta
_{\alpha }(\rho )-\delta _{\alpha }(\pi + \gamma + \rho ), \end{equation*}
and $\delta _{\alpha }(\pi ) = \delta (\pi )\overleftarrow{\partial }_{\alpha
}$.

Obviously, the following proposition takes place:

\proposition\label{alep} {\it $\aleph $ and $\aleph ^{\alpha }$ are
symmetrical and

1) $\aleph = 0$ for $n_- = 1$ and for $n_- = 2$,

2) $\aleph^{\alpha} = 0$ for $n_- = 0$, for $n_- = 2$, and for $n_- = 3$,

3) $\aleph \ne 0$ for $n_- \ge 4$ and $\aleph^\alpha \ne 0$ for $n_- \ge 4$.
}

This proposition allows us to find the partial solutions for the cases $n_- =
2,3$ and to prove that there is no solution for the case $n_- \ge 4$.

\subsection{Partial solution of Equation (\protect\ref{loceq}) for $\mathbf{
n_- = 2}$}

Due to Proposition~\ref{alep}, Eq.~(\ref{CE3}) reduced to $$[p,q]\Phi_1
(z,p,q,r) + \{{}F_1(z,q,r),zp\} + \mathrm{cycle}(p,q,r) = 0$$ for $n_- = 2$.
It has the solution ${}F_1(z,p,q) = 0$ and so $${}F(z,p,q) = c_{5}[\delta
(\pi )-\delta ( \gamma )].$$

\subsection{Partial solution of Equation (\protect\ref{loceq}) for $\mathbf{
n_- = 3}$} Eq. (\ref{CE3}) has the solution $${}F_1(z,p,q) = c_{5}(\delta
_{\alpha }(\pi) \gamma _{\alpha }-\delta _{\alpha }( \gamma )\pi _{\alpha
})$$ and so $${}F(z,p,q) = c_{5}(\delta (\pi )-\delta ( \gamma ) + \delta
_{\alpha }(\pi ) \gamma _{\alpha }-\delta _{\alpha }( \gamma )\pi _{\alpha
}).$$

\subsection{Solution of Equation (\protect\ref{loceq}) for $\mathbf{ n_- \ge
4}$} Here we prove that if $n_- \ge 4$ then Eq.~(\ref{loceq}) has a solution
only if $c_5 = 0$.

Let ${}F_1(z,p,q) = \sigma c_{5}[\delta _{\alpha }( \gamma )\pi _{\alpha
}-\delta _{\alpha }(\pi ) \gamma _{\alpha }] + {}F_2(z,p,q)$. Then
Eq.~(\ref{CE3}) acquires the form \begin{align} [p,q]\Phi_2 (z,p,q,r) +
\{{}F_2(z,q,r),zp\} + \mathrm{cycle}(p,q,r) = R(p,q,r), \label{CE6}
\end{align} where $$ R(p,q,r) = \sigma c_{5} \Big(\big([p,q] (\aleph - \aleph
\frac {\overleftarrow{\partial}}{\partial \rho_\alpha} \Big|_{\rho = 0}
\rho_\alpha) + \mbox{cycle}(p,q,r) \big) + 2\aleph ^{\alpha } \pi _{\alpha }
\gamma _{\alpha } \rho _{\alpha } \lambda_\alpha \Big) \in \oP_{2,2,2} $$.

\proposition\label{G} {\it Let $R(k,l,r)\in \oP_{2,2,2}$. Let the polynomial
$G(k,l) = -G(l,k)$ be the solution of the equation \begin{align} [k,l](G(k +
l,r)-G(k,r)-G(l,r)) + \{G(z,l,r),zk\} + \mathrm{cycle}(k,l,r) = R(k,l,r).
\label{CE7} \end{align} Then $G(k,l)\in \oP_{2,2}$ and does not depend on
$z$ up to coboundary, i.e. up to polynomials of the form \begin{equation}
\label{p-triv} M_{\mathrm{triv}}(z|k,l) = [k,l](t(z|k + l)-t(z|k)-t(z|l)) +
\{t(z|l),zk\}-\{t(z|k),zl\}. \end{equation} }

The proof of the Proposition based on the Propositions \ref{q4.2.5} -
\ref{q4.2.7} (see also \cite{n>2}).

Decompose the polynomial $G(z|k,l)$ as \begin{eqnarray} \label{decomp} G(k,l)
= t_0(z|k)-t_0(z|l) + t^{AB}_{11}(z)k_A l_B + t^A_1(z|k) l_A - t^A_1(z|l) k_A
+ \varphi (z|k,l), \end{eqnarray} where $t_0(z|k)\in \oP_1$, $t^A_1(z|k)\in
\oP_2$ and $\varphi(z|k,l)\in \oP_{2,2}$. The decomposition (\ref{decomp})
satisfies the following propositions:

\proposition\label{q4.2.5} {\it $t_0(z|k)-t_0(z|l) = dt(z|k,l) + [k,l]t(z)$,
where $t(z)\in {\cal A}$, and the linear form $t(z|f)$ is defined as $t(z|f)
= -t(z)f(z)$. }

To prove this proposition, we consider Eq.~(\ref{CE7}) for $r = 0$. We obtain
$[k,l](t_0(z|k + l)-t_0(z|k)-t_0(z|l))-\{t_0(z|k), z^Al_A\} + \{t_0(z|l),
z^Ak_A\} = 0$. This equation has the solution $t_0(z|k) = \alpha(1-\frac 1 2
z^Ak_A) + \{t(z),\,z^Ak_A\}$. It follows from $t_0(z|0) = 0$ that $\alpha =
0$, which implies the statement of the Proposition.

\proposition\label{q4.2.6} {\it The term $t_0(z|k)-t_0(z|l) +
t^{AB}_{11}(z)k_A l_B$ in~(\ref{decomp}) is a coboundary.}

Due to Proposition \ref{q4.2.5} we can write $t_0(z|k)-t_0(z|l) +
t^{AB}_{11}(z)k_A l_B = dt(z|k,l) + \tilde t^{AB}_{11}(z)k_A l_B$.

The terms linear in $k$, $l$ and $r$ in Eq.~(\ref{CE7}) give the equation
\begin{eqnarray} \sigma(C,A) \tilde t^{AB}_{11}(z)\DD^C + \sigma(A,B) \tilde
t^{BC}_{11}(z)\DD^A + \sigma(B,C) \tilde t^{CA}_{11}(z)\DD^B = 0.
\end{eqnarray} Together with $\tilde t^{AB}_{11}(z) = - \sigma(A,B)\tilde
t^{BA}_{11}(z)$, this equation implies that $ \sigma(A,B)\tilde
t^{AB}_{11}(z) = w^A(z) \DD^B - w^B(z)\DD^A$ with some vector-function
$w^A(z)$. So $\tilde t^{AB}_{11}(z)k_A l_B = \{w(z|k),zl\}-\{w(z|l),zk\}$
with $w(z|k) = w^A(z)k_A$. This coincides with (\ref{p-triv}).

\proposition\label{q4.2.7} {\it Up to a coboundary, $t^A_1(z|k)l_A -
t^A_1(z|l) k_A\in \oP_{2,2}$}.

Due to Propositions \ref{q4.2.5} and \ref{q4.2.6} we can assume that
$t_0(z|k)-t_0(z|l) + t^{AB}_{11}(z)k_A l_B = 0$. Consider the terms linear
in $l$ and $r$ in Eq. (\ref{CE7}). These terms give the equation:
$t^A_1(z|k)\DD^B - \sigma(A,B)t^B_1(z|k)\DD^A = 0 $, which implies
$t^A_1(z|k) = t(z|k)\DD^A$ and $t^A_1(z|k) l_A - t^A_1(z|l) k_A = \{t(z|k),
zl\}-\{t(z|l),zk\}$. So $t^A_1(z|k) l_A - t^A_1(z|l) k_A =
\{t(z|k),zl\}-\{t(z|l),zk\} -(t(z|k + l)-t(z|k)-t(z|l))[k,l] + (t(z|k +
l)-t(z|k)-t(z|l))[k,l] = d t(z|k,l) + \varphi_t (z|k,l)$, where $\varphi_t
(z|k,l) = (t(z|k + l)-t(z|k)-t(z|l))[k,l]\in \oP_{2,2}$. Thus, up to
coboundary, $G(k,l)\in \oP_{2,2}$.

Now, we can complete the proof of \ref{G} noticing that the linear in $r$
terms in Eq.~(\ref{CE7}) give $\{G(k,l),\, zr\} = 0$.

In such a way, ${}F_2(p,q)$ does not depend on $z$, and so ${}F_1(p,q)$ does
not depend on $z$ also, which implies that Eq.~(\ref{CE3}) acquires the form
\begin{align} & [p,q]\Phi_1 (p,q,r) + \mathrm{cycle}(p,q,r) = \label{CE4} \\
& \, = \sigma c_{5} \left( \aleph ([p,q] + [q,r] + [r,p]) + 2\aleph ^{\alpha
} \pi _{\alpha } \gamma _{\alpha } \rho _{\alpha } \lambda_\alpha\right)
\notag \end{align}

Let us note that the right hand side of (\ref{CE4}) change the sign under the
following transformations:

\begin{eqnarray} (\pi _{\alpha } \rightarrow -\pi _{\alpha },\ \gamma
_{\alpha } \rightarrow - \gamma _{\alpha },\ \rho _{\alpha } \rightarrow
-\rho _{\alpha }) \label{tr2} \end{eqnarray} for any $\alpha = 1,...,n_-$.
So, if Eq.~(\ref{CE4}) has a solution for $c_5 \ne 0$, then this equation has
a partial solution which change the sign under transformations (\ref{tr2}).
Let us look for such ${}F_1(p,q)$, that changes the sign under the
transformation (\ref{tr2}), i.e. for each $\alpha = 1,...,n_-$
\begin{eqnarray} {}F_1(a_1,a_2,\pi _{1},...,-\pi _{\alpha },...\pi _{n_{-}},
b_1,b_2, \gamma _{1},...,- \gamma_{\alpha },... \gamma _{n_{-}}) = \notag \\
= -{}F_1(a_1,a_2,\pi _{1},...,\pi _{\alpha },...\pi_{n_{-}}, b_1,b_2, \gamma
_{1},..., \gamma _{\alpha },... \gamma _{n_{-}}). \label{propo1}
\end{eqnarray}

\proposition\label{Proposition 2} \begin{equation*} \frac{\partial }{\partial
a_{i}}{}F_1(p,q) = \frac{\partial }{\partial b_{i}} {}F_1(p,q) = 0
\end{equation*}

Indeed, since $\deg {}F_1(p,q) = n_{-}$, the relation (\ref{propo1}) gives
the statement of the Proposition.

The decomposition of Eq.~(\ref{CE4}) on $a_i$, $b_i$ gives that this equation
reduces to the system of the equations \begin{eqnarray} \left\{
\begin{array}{l} {}F_1(p + q,r)-{}F_1(p,r)-{}F_1(q,r) = \sigma c_{5} \aleph
(\pi , \gamma ,\rho ) \\ c_{5} \aleph ^{\alpha } = 0 \label{alef0}
\end{array} \right.  \end{eqnarray}

Because $\aleph ^{\alpha } \neq 0$ for $n_{-} \ge 4$, the relation
(\ref{alef0}) implies, that $c_{5} = 0$ for $n_{-} \ge 4$.

Thus, the general solution of cohomology equation (\ref{5.4}) has the form

\begin{align} M_{2}(z|f,g) = &c_{1}m_{2|1}(z|f,g) + c_{2}m_{2|2}(z|f,g) +
c_{3}m_{2|3} (z|f,g) + \delta_{6,n_{-}}c_{4}m_{2|4}(z|f,g) + \nonumber\\ & +
c_{5}m_{2|5}(z|f,g) + d_{1}^{\mathrm{ad}} \zeta^{2}(z|f,g), \label{3.1a} \\
m_{2|1}(z|f,g) & = f(z)(\overleftarrow{\partial}_{A} \omega^{AB}
\partial_{B})^{3}g(z), \nonumber\\ m_{2|2}(z|f,g) & = \left\{ \begin{array}
[c]{l} \bar{f} \bar{g} \ \ \mbox{ for odd }n_- \\ 0\ \ \ \ \mbox{ for even
}n_-, \end{array} \right. \nonumber\\ m_{2|3}(z|f,g) & = \bar{f} \eE_z
g(z)-(-1)^{\varepsilon(f)\varepsilon(g)} \bar{g} \eE_z f(z) , \nonumber \\
m_{2|4}(z|f,g) & = \left\{ \begin{array} [c]{l} \int du[\{u^{A}
\partial_{A}f(u)\}g(u)-f(u)u^{A} \partial_{A}g(u)] \ \ \mbox{for }n_- = 6,\\
0\ \! \phantom{\int du[\{u^{A} \partial_{A}f(u)\}g(u)-f(u)u^{A}
\partial_{A}g(u)]} \mbox{for }n_- \ne 6, \end{array} \right. \nonumber\\
m_{2|5}(z|f,g) & = \left\{ \begin{array} [c]{l} L_2^{n_-} \phantom{0}
\;\;\;\;\;\;\;\;\;\;\;\;\;\;\;\;\;\;\;\;\;\; \mbox{for }0\le n_{-} \le 3,
\nonumber \\ 0 \phantom{L_2^{n_-}}
\;\;\;\;\;\;\;\;\;\;\;\;\;\;\;\;\;\;\;\;\;\; \mbox{for } n_{-} \geq 4,
\end{array} \right. \nonumber \end{align} where the bilinear forms $L_2^i$
are listed in Theorem \ref{th2} and $ d_{2}^{\mathrm{ad}}m_{2|i}(z|f,g,h) =
0,\;i = 1,...,5. $ Note that $\varepsilon(m_{2|5}) = n_{-}$.

\subsection{Independence and nontriviality}

It is shown in \cite{n>2} that no linear combination of nonzero $m_{2|i}$, $i
= 1,...,4$ can be expressed as coboundary, but zero.

Here we show that $m_{2|1}$, $m_{2|2}$, $m_{2|3}$, and $m_{2|5}$, represent
independent nontrivial cohomologies at $n_- = 1,3$ and that $m_{2|1}$,
$m_{2|3}$, and $m_{2|5}$, represent independent nontrivial cohomologies at
$n_- = 0,2$.

Suppose that $$ c_{1}m_{2|1}(z|f,g) + c_{2}m_{2|2}(z|f,g) +
c_{3}m_{2|3}(z|f,g) + c_{5} m_{2|5}(z|f,g) = d_{1}^{\mathrm{ad}}
\zeta(z|f,g). $$

Let the supports of the functions $f$ and $g$ satisfy the condition $$
z\bigcap\mathrm{supp}(f) = z\bigcap\mathrm{supp}(g) = \mathrm{supp}(f)\bigcap
\mathrm{supp}(g) = \varnothing. $$ In this case we obtain $c_{2} \bar{f}
\bar{g} = 0$, and $c_{2} = 0$.

{}Further, let $ \left[ z\bigcup\mathrm{supp}(f)\right]
\bigcap\mathrm{supp}(g) = \varnothing. $ Then we have $
c_{3}[(1-\frac{1}{2}z^{A} \frac{\partial}{\partial z^{A}})f(z)\bar{g}
-2c_{5}(-1)^{n_{-} \varepsilon(f)} \partial_{2}f(z) \Theta(z|g)
 = (-1)^{\varepsilon(f)\varepsilon_{M_{2}}}
\{f(z),\hat{\zeta}(z|g)\}. $ Considering the terms proportional to $f(z)$ in
this equation, we obtain $c_{3}f(z)\bar{g} = 0$, and $c_{3} = 0$. {}Further,
since $\varepsilon(M_{2}) = n_{-}$ we have $-2c_{5}
\partial_{2}f(z)\Theta(z|g) = \{f(z),\hat {\zeta}(z|g)\}$ and
\begin{equation} \partial_{2} \hat{\zeta}(z|u) = 0,\label{3.2} \end{equation}
\begin{equation} 2c_{5} \delta(x_{1}-y_{1})\theta(x_{2}-y_{2}) = \partial_{1}
\hat{\zeta }(z|u).\label{3.3} \end{equation} It follows from
Eq.~(\ref{3.2}) that $$ \zeta(z|g) = \sum_{l,m \geq
0}V_{1}^{lm}(\xi|u)\partial_{1}^{l} \partial_{2}^{m} \delta(x-y) + \sum_{l
\geq 0}V_{2}^{l}(\xi|u)\partial_{1}^{l} \delta
(x_{1}-y_{1})\theta(x_{2}-y_{2}) + \mu(x_{1},\xi|u). $$ Then we obtain from
Eq.~(\ref{3.3}) $$ \left( 2c_{5} \delta(x_{1}-y_{1})- \sum_{l \geq
0}V_{2}^{l}(\xi|u)\partial _{1}^{l + 1} \delta(x_{1}-y_{1})\right)
\theta(x_{2}-y_{2}) = \partial_{1} \hat {\mu}(x_{1},\xi|u) $$ which implies
$2c_{5} \delta(x_{1}-y_{1}) = \sum_{l \geq 0}V_{2}^{l}(\xi|u)\partial_{1} ^{l
+ 1} \delta(x_{1}-y_{1})$ and $c_{5} = 0$.

It is proved in \cite{n>2} that $m_{2|1}$ is not coboundary, and so $c_1 = 0$
also.

\subsection{Exactness of the form $d_1^{\rm ad} \zeta(z|f,g)$}

Let us discuss the terms $M_{d|2}(z|f,g)$, \begin{eqnarray*}
M_{d|2}(z|f,g)\equiv d_1^{\rm ad} \zeta^2(z|f,g) =
\sigma(f,M_2)\{f(z),\zeta^2(z|g)\}- \\ - \sigma(f,g)
\sigma(g,M_2)\{g(z),\zeta^2(z|f)\}- \zeta^2(z|\{f,g\}) \end{eqnarray*} in the
expression (\ref{3.1a}).

Assume that both functions $M_2(z|f,g)$ and $\zeta^2(z|g)$ belong to the
space ${\cal A}$ for all functions $f,\,g\in \mathbf D^{n_-}_{2}$.

1) ${\cal A} = \mathbf D^{\prime n_-}_{2}$. In this case $\zeta^2(z|g)\in
\mathbf D^{\prime n_-}_{2}$ and the form $M_{d|2}(z|f,g)$ is exact (trivial
cocycle).

2) ${\cal A} = \mathbf E^{n_-}_{2}$. In this case $\zeta^2(z|g)\in \mathbf
E^{n_-}_{2}$ and the form $M_{d|2}(z|f,g)$ is exact (trivial cocycle).

3) ${\cal A} = \mathbf D^{n_-}_{2}$.

Consider the condition that $M_2(z|f,g)\in \mathbf D^{n_-}_{2}$ for all
$f,g\in\mathbf D^{n_-}_{2}$. This condition gives that $c_2 = c_4 = 0$ in
(\ref{3.1a}) and that $N^E_2(z|f,g) + d_1^{\rm ad} \zeta^2(z|f,g)\in \mathbf
D^{n_-}_{2}$. Let $\zeta^2(z|f) = N_1(z|f) + \zeta(z|f) $, where $N_1(f) =
-2\Lambda(x_2)\int du \theta(x_1-y_1)f(u),$ and $\Lambda\in C^\infty({\mathbb
R})$ be a function of $x$ such that $\frac d {dx} \Lambda \in \D(\R)$ and
$\Lambda(-\infty) = 0,\ \Lambda( + \infty) = 1$.

Then $d_1^{\mathrm{ad}}N_1(f,g) = -\Lambda (x_{2})\int du\delta
(x_{1}-y_{1})[\partial _{2}f(u)g(u)-f(u)\partial _{2}g(u)]$ and $N^D_2(z|f,g)
\stackrel {def} = N^E_2(f,g) + d_1^{\mathrm{ad}}N_1(f,g) \in \mathbf
D^{n_-}_{2}$ for all $f,g\in\mathbf D^{n_-}_{2}$ because
$\Theta(z|\partial_{2}fg)-\Theta(z|f\partial_{2}g)- \Lambda (x_{2})\int
du\delta (x_{1}-y_{1})[\partial _{2}f(u)g(u)-f(u)\partial _{2}g(u)]\in
\mathbf D^{n_-}_{2}$.

{}Further, let us use the following Lemma proved in \cite{n>2}:

\lemma\label{11111}{\it Let $d_1^{\rm ad} \zeta(z|f,g)\in \mathbf D^{n_-}_{n_
+ }$. Then \begin{eqnarray} \label{lrepr} \zeta(z|f) = \zeta_D (z|f) +
\zeta(z)\bar{f}, \end{eqnarray} where $\zeta_D(z|f) \in {\mathbf D}^{n_-}_{n_
+ }$ and $\zeta(z)\in \mathbf E^{n_-}_{n_ + }$. }

Thus, $ M_{d|2}(z|f,g) = M_{d|2|D}(z|f,g) + M_{d|2|C}(z|f,g), $ where $
M_{d|2|D}(z|f,g) = d_1^{\rm ad} \zeta_D(z|f,g) $ are the exact forms, and $
M_{d|2|C}(z|f,g) = \{f(z),\zeta(z)\} \bar{g}- \sigma(f,g)\{\zeta(z),g(z)\}
\bar{f} $ are nontrivial cocycles parameterized by the classes of the
functions ${\raise2pt\hbox{$\mathbf E^{n_-}_{2}$}}
\big/{\raise-2pt\hbox{$\mathbf D^{n_-}_{2}$}}$.

{}Finally, we have 
\begin{eqnarray}
 M_{2}(z|f,g) &=& c_{1}m_{2|1}(z|f,g) + c_{3}m_{2|3}(z|f,g)
+ c_5 (m_{2|5}(z|f,g) + 
d_{1}^{\mathrm{ad}}N_1 (z|f,g)) + \nonumber\\
&+&
\{f(z),\zeta^E(z)\} \bar{g}- \sigma(f,g)\{\zeta^E(z),g(z)\}
\bar{f} + d_{1}^{\mathrm{ad}}\zeta^D(z|f,g), 
\nonumber
\end{eqnarray}
 where
$\zeta^E(z)\in \mathbf E^{n_-}_{2}/\mathbf D^{n_-}_{2}$.


\setcounter{equation}{0} \def\theequation{A\arabic{appen}.\arabic{equation}}

\newcounter{appen} \newcommand{\appen}[1]{\par\refstepcounter{appen}
{\par\medskip\noindent\Large\bf Appendix \arabic{appen}. \medskip }{\bf
\large{#1}}}

\renewcommand{\theorem}{\par\refstepcounter{theorem} {\bf Theorem
A\arabic{appen}.\arabic{theorem}. }}
\renewcommand{\lemma}{\par\refstepcounter{lemma} {\bf Lemma
A\arabic{appen}.\arabic{lemma}. }}
\renewcommand{\proposition}{\par\refstepcounter{proposition} {\bf Proposition
A\arabic{appen}.\arabic{proposition}. }} \makeatletter
\@addtoreset{theorem}{appen} \makeatletter \@addtoreset{lemma}{appen}
\makeatletter \@addtoreset{proposition}{appen}
\renewcommand\thetheorem{A\theappen.\arabic{theorem}}
\renewcommand\thelemma{A\theappen.\arabic{lemma}}
\renewcommand\theproposition{A\theappen.\arabic{proposition}}

\renewcommand{\subsection}[1]{\refstepcounter{subsection} {\bf
A\arabic{appen}.\arabic{subsection}. }{\ \bf #1}}
\renewcommand\thesubsection{A\theappen.\arabic{subsection}} \makeatletter
\@addtoreset{subsection}{appen}

\renewcommand{\subsubsection}{\par\refstepcounter{subsubsection} {\bf
A\arabic{appen}.\arabic{subsection}.\arabic{subsubsection}. }}
\renewcommand\thesubsubsection{A\theappen.\arabic{subsection}.\arabic{subsubsection}}
\makeatletter \@addtoreset{subsubsection}{subsection}

\newcommand {\subsubsub}{\par\refstepcounter{subsubsub}{\bf
A\arabic{appen}.\arabic{subsection}.\arabic{subsubsection}.\arabic{subsubsub}.}}

\renewcommand\thesubsubsub{A\theappen.\arabic{subsection}.\arabic{subsubsection}.\arabic{subsubsub}}
\makeatletter \@addtoreset{subsubsub}{subsubsection}

\setcounter{equation}{0}

\newpage
\appen {} \label{App3}

In this appendix the lemmas, which are used in the proof of the Theorems
\ref{th4.3.1} and \ref{th4.3.2}, are formulated. These lemmas are proved in
\cite{n>2}.

Let $p$, $q$ and $r$ be 3 sets of supervariables with $2$ even and $n_-$ odd
variables each. The notation $[p,q]$ in this appendix means as before $[p,q]
= (-1)^A p_A \omega^{AB} q_B$.

\lemma\label{lA1.1} {\it Let $p$ be supervariables with $n \geq 2$ even and
with no odd variables. Let ${}F(p)$, $G(p)$, $u(p)$ and $v(p)$ be the
polynomials in $p$ such that $u(p)$ and $v(p)$ does not depend on $p_1$. Let
$v|_{p_2 = 0} \neq 0$. Then the following relation \begin{eqnarray}
\label{LA3.3eq} (p_1 p_2 + u) {}F + v G = 0 \end{eqnarray} is satisfied if
and only if there exists such polynomial $f(p)$ that \begin{eqnarray}
\label{LA3.3sol} {}F = v f(p),\ \ \ G = -(p_1 p_2 + u)f(p).  \end{eqnarray} }

\lemma\label{lA1.2} {\it Let $n_ + \geq 2$, $n_- \geq 0$. If ${}F(p,q,r)$ and
$G(p,q,r)$ are polynomial in $p$, $q$ and $r$ then the following relation
\begin{eqnarray} \label{LA3.1eq} [p,q]{}F + [q,r]G = 0 \end{eqnarray} is
satisfied if and only if there exists such polynomial $f(p,q,r)$ that
\begin{eqnarray} \label{LA3.1sol} {}F = [q,r]f(p,q,r),\ \ \ G =
-[p,q]f(p,q,r).  \end{eqnarray} }

\lemma\label{lA3} {\it If $n_ + \ge2$ then the general solution of the
equation \begin{equation} \frac{\partial f(k,l)}{\partial k_A} =
l^Ag(k,l),\mbox{ where } l^A = (-1)^{\varepsilon_A} \omega^{AB}l_B
\label{A4.1} \end{equation} for the polynomial functions $f(k,l)$ and
$g(k,l)$ has the form $$ f(k,l) = H(x,l) = \sum_{q = 0}^Qf_q(l)x^q,\quad
g(k,l) = \frac{\partial H(x,l)}{\partial x}, $$ where $f_q(l)$ are some
polynomials of $l$, and $x = k_Al^A$.}

\lemma\label{lA4} {\it If $n_ + \neq 1$ then the general solution of the
equation \begin{equation} k_Aa_B(k)- \sigma(A,B)k_Ba_A(k) = 0, \label{A6.1}
\end{equation} for the polynomials $a_A(k)$ has the form $$ a_A(k) = k_Aa(k),
$$ where $a(k)$ is an arbitrary polynomial.}

\setcounter{equation}{0}

\newpage
\appen{Homogeneous equation for the local cocycles} \label{hom}

In this Appendix we find the solution of the homogeneous equation
\begin{equation} \label{general} [p,q]\Psi (z|p,q,r) + \{\psi (z|p,q),zr\} +
\mathrm{cycle}(p,q,r) = 0, \end{equation} where $\Psi (z|p,q,r) = \psi (z|p +
q,r)-\psi (z|p,r)-\psi (z|q,r)$, and $\psi(z|p,q) = -\psi(z|q,p)$ is
polynomial on $p$ and $q$. Due to antisymmetry, $\psi(z|0,0) = 0$.

It follows from Propositions \ref{G}, that up to coboundary the polynomial
$\psi(z|k,l)\in \oP_{2,2}$ and does not depend on $z$.

In such a way the homogeneous part of Eq.~(\ref{general}) for the local
bilinear form reduces to \begin{equation} [k,l]\Psi(k,l,r) + [l,r]\Psi(l,r,k)
+ [r,k]\Psi(r,k,l) = 0, \label{5.12} \end{equation} \begin{equation}
\Psi(k,l,r) = \psi(k + l,r)-\psi(k,r)-\psi(l,r), \label{5.12a} \end{equation}
where $\psi(k,l) = -\psi(l,k)$ and $\psi\in \oP_{2,2}$, i.e. $ \psi(k,l) =
\sum_{p,q = 2}^Nm^{(A)_p|(B)_q}(k_A)^p(l_B)^q. $

\subsection{Solution of the equation (\ref{5.12}). The case $n_- = 0.$}

\theorem\label{th4.3.1} {\it Let $n_ + = 2$, $n_- = 0$. Let $\psi(p,q) =
-\psi(q,p)$ be a polynomial such that $\psi\in \oP_{2,2}$, let $[p,q]
\stackrel {def} = p_{i} \omega ^{ij}q_{j}$ and let \begin{equation}
\Psi(p,q,r) = \psi(p + q,r)-\psi(p,r)-\psi(q,r), \label{4.3.1a}.
\end{equation} Then the general solution of the equation \begin{equation}
[p,q]\Psi(p,q,r) + [q,r]\Psi(q,r,p) + [r,p]\Psi(r,p,q) = 0, \label{eq}
\end{equation} has the form \begin{equation} \psi(p,q) = c([p,q])^3 +
[p,q]\bigl(\varphi(p + q)-\varphi(p)-\varphi(q)\bigr), \label{4.3.1b}
\end{equation} where $c$ is an arbitrary constant, $\varphi(p)$ is an
arbitrary polynomial satisfying the condition $\varphi(0) = 0$.}

{}For any pair $a_i$, $i = 1,2$ define $a^1 = a_2$, $a^2 = -a_1$, and for any
two pairs $a_i$, $b_i$, $i = 1,2$ we define $[a,b] = -[b,a] = a_1b_2-a_2b_1 =
a_ib^i$.

Let us introduce the notation: $$x = [k,l],\quad y = [l,r],\quad z = [r,k].$$

It is easily to check the correctness of the following identities
\begin{equation} r_i [k,l] + k_i[l,r] + l_i[r,k]\equiv0. \label{A3.9}
\end{equation}

Before proving the Theorem \ref{th4.3.1} let us prove the following 3
propositions

\proposition\label{q4.3.1.1} {\it Let $k$, $l$ and $r$ be the supervariables
with 2 even variables and with no odd variables. If ${}F(k,l,r)$, $G(k,l,r)$
and $H(k,l,r)$ are polynomial in $k$, $l$ and $r$ then the following relation
\begin{eqnarray} \label{A3.3} [k,l]{}F + [l,r]G + [r,k]H = 0 \end{eqnarray}
is satisfied if and only if there exist such polynomials $s_i(k,l,r)$, ($i =
1,2$) that \begin{equation} {}F = [r,s] = r_1 s_2-r_2 s_1,\quad G = [k,s],
\quad H = [l, s], \label{A3.10} \end{equation} }

{\it Proof.} Multiplying the relation (\ref{A3.3}) by $k_i$ and using the
identities (\ref{A3.9}) we obtain $$ x(k_i{}F-r_iG) + z(k_iH-l_iG) = 0, $$
which gives according to the Lemma \ref{lA1.1} that $$ k_i{}F-r_iG = z
s_i,\quad k_iH-q_iG = -x s_i, $$ where $s_{i}$ are some polynomials.
Multiplying these relations by $k^i$, $l^i$ or $r^i$ we obtain the
expressions (\ref{A3.10}).  Besides, it is easily to check that the
expressions (\ref{A3.10}) satisfy Eq.~(\ref{A3.3}) for arbitrary polynomials
$s_i$, q.e.d.

In fact, the functions ${}F$, $G$ and $H$ in Eq.~(\ref{A3.3}) are connected
by the relations $G(k,l,r) = {}F(l,r,k)$, $H(k,l,r) = {}F(r,k,l)$, and,
besides, ${}F(l,k,r) = {}F(k,l,r)$.

\proposition\label{q4.3.1.2} {\it Let $n_ + = 2$, $n_- = 0$. Then the general
solution of the equation \begin{equation} x\Psi(k,l,r) + y\Psi(l,r,k) +
z\Psi(r,k,l) = 0 \label{A3.11} \end{equation} for the polynomial $\Psi(k,l,r)
= \Psi(l,k,r)$ has the form \begin{equation} \label{phi} \Psi = [r,\phi],
\end{equation} where $\phi_i$ are arbitrary symmetrical polynomials of the
variables $k$, $l$ and $r$.}

{\it Proof.} Using the Proposition \ref{q4.3.1.1} we can write $$ \Psi(k,l,r)
= [r,s(k,l,r)], \Psi(l,r,k) = [k,s(k,l,r)], \Psi(r,k,l) = [l,s(k,l,r)], $$
which implies $\Psi(k,l,r) = [r,s'(k,l,r)]$, where $ s'_i(k,l,r) =
\frac{1}{3}[s_i(k,l,r) + s_i(l,r,k) + s_i(r,k,l)] $ are cyclically
symmetrical polynomials. Taking into account the symmetry property of the
function $\Psi$ we obtain $ \Psi(k,l,r) = [r,\phi(k,l,r)], $ where $
\phi_i(k,l,r) = \frac{1}{2}[s'_i(k,l,r) + s'_i(l,k,r)] $ are symmetrical
polynomials, q.e.d.

\proposition\label{q4.3.1.3} {\it If $n_ + = 2$ and $n_- = 0$ then the
solution of Eqs.~(\ref{4.3.1a})-(\ref{eq}) has the form \be \label{prop}
\psi(p,q) = [p,q]g(p,q), \ee where $g$ is some polynomial such that $g(p,q) =
g(q,p)$ and $g(p,q)\in \oP_{1,1}$}.

{\it Proof.} Consider the representations (\ref{4.3.1a}) and (\ref{phi}) for
the function $\Psi$. We have $ \psi(k + l,r)-\psi(k,r)-\psi(l,r) =
r_i\phi^i(k,l,r) $ Apply the operator $\partial/\partial l_j\biggr|_{l = 0}$
to this relation: $ \frac{\partial\psi(k,r)}{\partial k_j} =
r_i\phi^{ij}(k,r)$, where $ \phi^{ij}(k,r) =
\frac{\partial\phi^i(k,l,r)}{\partial l^j} \biggr|_{l = 0}$. Due to symmetry
of the functions $\phi^i$, $ \phi^{ij}(k,r) = \phi^{ij}(r,k). $ {}From the
antisymmetry of $\psi$ it follows also that $
\frac{\partial\psi(k,r)}{\partial r_j} = -k_i\phi^{ij}(k,r), $ and we find $
N_{kr} \psi(k,r) = (k_ir_j-k_jr_i)\phi^{ij}(k,r) = [k,r]g'(k,r), $ where $
N_{kr} = k_i\frac{\partial}{\partial k_i} + r_i\frac{\partial}{\partial
r_i}$, $ g'(k,r) = \phi^{12}(k,r)-\phi^{21}(k,r) = g'(r,k), $ which implies $
\psi(k,l) = [k,l]g(k,l). $ Obviously, $g(k,r) = g(r,k)$ and $g\in\oP_{1,1}$,
q.e.d.

\proposition\label{q4.3.1.4} \begin{eqnarray} \label{prop2} g(p,q) =
\varphi(p + q)-\varphi(p)-\varphi(q) + {}F(\left [p,q\right ]^2),
\end{eqnarray} {\it where $\varphi(p)$ and ${}F$ are some polynomials such
that $\varphi \in\oP_1$.}

Let us substitute (\ref{prop}) into (\ref{eq}). One can reduce the result to
the following form:

\begin{eqnarray} \left [p,q\right ](\left [q,r\right ]-\left [r,p\right
])W(p,q,r) = \left [r,p\right ](\left [p,q\right ]-\left [q,r\right
])W(q,r,p), \end{eqnarray} where $W(p,q,r) = g(p + q,r)-g(q + r,p) +
g(p,q)-g(q,r)$

Using Lemma \ref{lA1.1} 2 times we can write \begin{eqnarray} W(p,q,r) =
\left [r,p\right ]\left [r + p,q\right ]U(p,q,r) \end{eqnarray} with some
polynomial $U$, such that $U(p,q,r) = U(q,r,p)$.

It follows from $W(r,q,p) = -W(p,q,r)$ that $U(r,q,p) = U(p,q,r)$ and $U$ is
symmetrical polynomial.

Let us apply the operator $\partial_{r_A}|_{r = 0}$ to the formula
\begin{eqnarray} \label{formula} g(p + q,r)-g(q + r,p) + g(p,q)-g(q,r) =
\left [r,p\right ]\left [r + p,q\right ]U(p,q,r). \end{eqnarray} We obtain
$\varphi^A(p + q)-\partial_{q_A} g(p,q) - \varphi^A(q) = p^A \left [p,q\right
] U(p,q,0)$, where $\varphi^A(q) = \partial_{r_A}g(q,r)|_{r = 0}$ and finally
\begin{eqnarray} \label{der} \partial_{q_A} g(p,q) = \varphi^A(p + q) -
\varphi^A(q) - p^A \left [p,q\right ] U(p,q,0), \end{eqnarray} The conditions
$\partial_{q_B} \partial_{q_A} - \sigma (A,B) \partial_{q_A} \partial_{q_B} =
0$ leads to \begin{eqnarray} \!\!\!\!\!\!\!\!\!\!\!\!\!\!\!\!\!
\partial_{q_B} \varphi^A(p + q) - \partial_{q_B} \varphi^A(q) + \sigma(A,B)
p^A p^B U(p,q,0) - \sigma (A,B) p^A \left [p,q\right ] \partial_{q_B}
U(p,q,0) = \nonumber \\ \left( \sigma (A,B) \partial_{q_A} \varphi^B(p + q) -
\sigma (A,B)\partial_{q_A} \varphi^B(q) + p^B p^A U(p,q,0) - p^B \left
[p,q\right ] \partial_{q_A} U(p,q,0) \right) \end{eqnarray} so
\begin{eqnarray} \label{eq2} \partial_{q_B}(\varphi^A(p + q) - \varphi^A(q))-
\sigma (A,B) \partial_{q_A}(\varphi^B(p + q) - \varphi^B(q)) = \nonumber \\
\left [p,q\right ] \sigma (A,B) ( p^A \partial_{q_B} - \sigma (A,B) p^B
\partial_{q_A}) U(p,q,0) \end{eqnarray} To solve (\ref{eq2}) consider the
case $q = 0$. Then (\ref{eq2}) gives $ \partial_{p_B} \varphi^A(p) - \sigma
(A,B) \partial_{p_A} \varphi^B(p) = C_{BA} - \sigma (A,B)C_{AB} $, where
$C_{AB} = \partial_{p_A} \varphi^B(p)|_{p = 0} = \partial_{p_A}
\partial_{q_B}g(p,q)|_{p = q = 0} = \sigma(A,B) C_{BA}$ because $g(p,q) =
g(q,p)$.

So $ \partial_{p_B} \varphi^A(p) - \sigma (A,B) \partial_{p_A} \varphi^B(p) =
0 $ and $ \varphi^A(p) = \partial_{p_A} \varphi(p). $ with some polynomial
$\varphi$. Eq.~(\ref{eq2}) now has the form: \begin{eqnarray} \label{pseudo}
(p^A \partial_{q_B} - \sigma (A,B) p^B \partial_{q_A}) U(p,q,0) = 0
\end{eqnarray} It follows from (\ref{pseudo}) by Lemma \ref{lA4} that
$\partial_{q_A}U(p,q,0) = p^A V(p,q)$ with some polynomial $V(p,q)$ and then
by Lemma \ref{lA3} that \begin{eqnarray} U(p,q,0) = H(\left [p,q\right ])
\end{eqnarray} with some polynomial $H$.

Now (\ref{der}) have the form $ \partial_{q_A} g(p,q) = \partial_{q_A}
\varphi(p + q) - \partial_{q_A} \varphi(q) - p^A \left [p,q\right ] H(\left
[p,q\right ]) $ and hence $ g(p,q) = \varphi(p + q) - \varphi(q) -\varphi(p)
- K(\left [p,q\right ]) + L(p) $ with some polynomials $K$ and $L$. The
condition $g(p,0) = 0$ gives $L = 0$, and $g(p,q) = g(q,p)$ gives $K =
{}F(\left [p,q\right ]^2)$. q.e.d.

{\it Proof of the Theorem \ref{th4.3.1}.} Now we know that the solution of
(\ref{eq}) has the form \begin{eqnarray} \psi(p,q) = \left [p,q\right
]{}F(\left [p,q\right ]^2) + \psi_1(p,q), \end{eqnarray} where
\begin{eqnarray} \psi_1(p,q) = \left [p,q\right ](\varphi(p + q) - \varphi(q)
-\varphi(p)) \end{eqnarray} The polynomial $\psi_1$ satisfies Eq.~(\ref{eq})
for arbitrary polynomial $\varphi\in\oP_1$. So $\left [p,q\right ]{}F(\left
[p,q\right ]^2)$ has to satisfy this equation also.

Evidently, polynomial $\psi = \left [p,q\right ]{}F(\left [p,q\right ]^2)$
should satisfy (\ref{eq}) in every power on $\left [p,q\right ]$.

It is easy to check that if $\psi = [p,q]^{2n + 1}$ satisfies (\ref{eq}) then
$n = 1$. q.e.d.

\subsection{The solution of the equation (\ref{5.12}). The case $n_- \geq
0$.}

\theorem\label{th4.3.2} {\it Let $n_ + = 2$, $n_- \geq 0$. Let $\psi(p,q) =
-\psi(q,p)$ be a polynomial such that $\psi\in \oP_{2,2}$, let $[p,q]
\stackrel {def} = (-1)^{\varepsilon _{A}}p_{A} \omega ^{AB}q_{B}$ and let
\begin{equation} \Psi(p,q,r) = \psi(p + q,r)-\psi(p,r)-\psi(q,r),
\label{4.3.2a}.  \end{equation} Then the general solution of the equation
\begin{equation} [p,q]\Psi(p,q,r) + [q,r]\Psi(q,r,p) + [r,p]\Psi(r,p,q) = 0,
\label{4.3.2eq} \end{equation} has the form \begin{equation} \psi(p,q) =
c([p,q])^3 + [p,q]\bigl(\varphi(p + q)-\varphi(p)-\varphi(q)\bigr),
\label{4.3.2sol} \end{equation} where $c$ is an arbitrary constant,
$\varphi(p)$ is an arbitrary polynomial satisfying the condition $\varphi \in
\oP_1 $.}

{\it Proof.} We solve the theorem by induction on the number of odd
generating elements $n_-$. Namely, below we show that the general solution of
Eq.~(\ref{4.3.2eq}) has the form (\ref{4.3.2sol}), where $ \varepsilon(c) =
\varepsilon(\varphi) = \varepsilon(M_2). $

{}First of all let us note that the expression (\ref{4.3.2sol}) does be the
solution of Eq.~(\ref{4.3.2eq}) at arbitrary $n_-$. Let us assume, that the
expression (\ref{4.3.2sol}) is the general solution of Eq.~(\ref{4.3.2eq}) at
$n_-\le N$ (as it stated by Theorem \ref{th4.3.1}, this assumption is true at
$N = 0$). Let us find the solution of this equation at $n_- = N + 1$.

Let us introduce some notation.

Let $K$, $L$ and $R$ denote some sets of $2$ even variables and $N + 1$ odd
variables, $$ K = (k,\mu),\quad L = (l, \nu),\quad R = (r, \sigma),\quad
\varepsilon(\mu) = \varepsilon( \nu) = \varepsilon( \sigma) = 1, $$ where
$k$, $l$, $r$ are the sets of $2$ even variables and the first $N$ odd
variables, while $\mu$, $ \nu$, $ \sigma$ are the last ($N + 1$)-th odd
variables. {}Further, let $$ x = [k,l],\quad y = [l,r],\quad z = [r,k], $$ $$
X = [K,L] = x + \lambda\mu \nu,\ \ Y = [L,R] = y + \lambda \nu \sigma,\ \ Z =
[R,K] = z + \lambda \sigma\mu, \ \ \lambda = \omega^{2 + n_-,2 + n_-} = \pm1.
$$

So, let the general solution of (\ref{4.3.2eq}) for $n_- = N$ be $
\psi_{(N)}(k,l) = c([k,l])^3 + [k,l]\bigl( \varphi_{(N)}(k +
l)-\varphi_{(N)}(k)-\varphi_{(N)}(l)\bigr), $ $ \varphi_{(N)}(0) = 0, $ where
the subscript in the round brackets denotes the number of odd arguments.

Obviously, $\psi_{(N + 1)}(K,L)|_{\mu = \nu = \sigma = 0} = \psi_{(N)}(k,l)$.
So, we can rewrite the function $\psi_{(N + 1)}(K,L)$ in the form
\begin{equation} \psi_{(N + 1)}(K,L) = \psi_0(K,L) + \mu\psi_1(k,l)-
\nu\psi_1(l,k) + \mu \nu\psi_{12}(k,l), \label{513q} \end{equation} where
$\psi_0(K,L) = c([K,L])^3 + [K,L]\bigl( \varphi_{(N)}(k +
l)-\varphi_{(N)}(k)-\varphi_{(N)}(l)\bigr)$, and the functions $\psi_1(k,l)$
and $\psi_{12}(k,l)$ possess the properties $\psi_1(k,l)\in \oP_{1,2}$,
$\psi_{12}(k,l) = \psi_{12}(l,k)\in \oP_{1,1}$ following from the properties
of the function $\psi_{(N + 1)}(K,L)\in \oP_{2,2}$. Substitute the
representation (\ref{4.3.2sol}) of the function $\psi_{(N + 1)}(K,L)$ in the
equation \begin{equation} X\Psi_{(N + 1)}(K,L,R) + Y\Psi_{(N + 1)}(L,R,K) +
Z\Psi_{(N + 1)}(R,K,L) = 0.  \label{5.14} \end{equation} The independent
equations arise from the terms of (\ref{5.14}) proportional to $\mu \nu$,
\begin{equation} y[\psi_{12}(l + r,k)-\psi_{12}(l,k)] = z[\psi_{12}(r +
k,l)-\psi_{12}(k,l)], \label{5.15} \end{equation} (and cyclic $k\rightarrow
l\rightarrow r\rightarrow k$), and from the terms proportional to $\mu$,
\begin{equation} x[\psi_1(k + l,r)-\psi_1(k,r)] + z[\psi_1(r +
k,l)-\psi_1(k,l)] = y[\psi_1(k,l + r)-\psi_1(k,l)-\psi_1(k,r)], \label{5.16}
\end{equation} (and cyclic $k\rightarrow l\rightarrow r\rightarrow k$). The
equation arising from the monomials proportional to $\mu \nu \sigma$ turns
out to be consequence of two former equations.

At first consider Eq.~(\ref{5.15}).

It follows from it by Lemma \ref{lA1.2} that $ \psi_{12}(r +
k,l)-\psi_{12}(k,l) = [l,r]a(k,l,r), $ where $a(k,l,r)$ is a polynomial. Let
us apply the operator $\partial/\partial_{r_A}|_{r = 0}$ to this polynomial:
\begin{equation} \frac{\partial \psi_{12}(k,l)}{\partial k^A} =
l^Aa(k,l),\quad l^A = (-1)^{\varepsilon_A} \omega^{AB}l_B,\quad a(k,l) =
-a(k,l,0). \label{5.17} \end{equation} The general solution of (\ref{5.17})
is (see Lemma~\ref{lA3}) $ \psi_{12}(k,l) = h(x), $ where $h(x)$ is some even
polynomial of $x = [k,l]$. Let the highest order of the polynomial $h(x)$ be
equal to $2p$. Consider the ($4p + 2$)-order terms in Eq.~(\ref{5.15}): $
y[(x-z)^{2p}-x^{2p}] = z[(x-y)^{2p}-x^{2p}]$. This equation for $p$ has the
unique solution $p = 0$, which implies $\psi_{12}(k,l) = {\rm const}$. Due to
the requirement $ \psi_{12}(0,0) = 0$ we obtain finally $
\psi_{12}(k,l)\equiv 0. $

Consider Eq.~(\ref{5.16}). Let us apply the operator $\partial/\partial
k_A|_{k = 0}$ to it: \begin{equation} l^A\psi_1(l,r)-r^A\psi_1(r,l) = y[a^A(l
+ r)-a^A(l)-a^A(r)], \label{5.19} \end{equation} \begin{equation} \mbox{where
} a^A(l) = \frac{\partial \psi_1(k,l)}{\partial k_A} \biggr|_{k = 0} \in
\oP_2. \label{5.20} \end{equation} Multiplying Eq.~(\ref{5.19}) by $r_A$, we
obtain $ \psi_1(l,r) = -r_D[a^D(l + r)-a^D(l)-a^D(r)]. $ Applying the
operator $\partial/\partial l_A|_{l = 0}$ to this equation we obtain $$
\frac{\partial \psi_1(l,r)}{\partial l_A} \biggr|_{l = 0} = a^A(r) = -
\sigma(A,D)r_D\frac{\partial a^D(r)}{\partial r_A} = -
\frac{\partial[r_Da^D(r)]}{\partial r_A} + a^A(r), $$ which implies
\begin{equation} r_Aa^A(r) = 0,\quad \psi_1(l,r) = -r_A[a^A(l + r)-a^A(l)].
\label{5.21} \end{equation} Substituting the relation (\ref{5.21}) in the
left-hand part of Eq.~(\ref{5.19}), applying the operator
$\partial^2/\partial l_B\partial l_C\bigl|_{l = 0}$ to the resulting
equation, and taking into account the properties (\ref{5.20}), we obtain
\begin{eqnarray*} \sigma(A,B) \sigma(A,C)r^A\frac{\partial a^B(r)}{\partial
r_C} + \sigma(A,B) \sigma(A,C) \sigma(B,C)r^A\frac{\partial a^C(r)}{\partial
r_B} + \omega^{AB}a^C(r) + \\ + \sigma(A,B)\omega^{A,C}a^B(r) =
r^C\frac{\partial a^A(r)}{\partial r_B} + \sigma(B,C)r^B\frac{\partial
a^B(r)}{\partial r_C}. \end{eqnarray*} Multiplying this equation by $r_C$,
we obtain $ \sigma(A,B)r^AN_ra^B(r) = r^BN_ra^A(r), $ where $ N_r =
r_D\frac{\partial}{\partial r_D}, $ is Euler operator, or \begin{equation}
r^Aa^B(r)- \sigma(A,B)r^Ba^A(r) = 0, \label{5.21a}, \end{equation} which in
view of (\ref{5.20}) implies (see Lemma~\ref{lA4}) that $ a^A(r) =
r^Aa(r),\quad a(r)\in \oP_1, $ which gives $ \psi_1(k,l) = x[a(k + l)-a(k)].
$ {}Finally, we have $$ \psi_{N + 1}(K,L) = c\biggl([K,L]\biggr)^3 +
[K,L]\biggl(\varphi_{N + 1}(K + L)-\varphi_{N + 1}(K)-\varphi_{N +
1}(L)\biggr), $$ $$ \varphi_{N + 1}(K) = \varphi_{N}(k) + \mu a(k). $$

Thus, the general solution $\psi(k,l)$ of Eq.~(\ref{5.12}) indeed has the
form (\ref{5.13}),

\setcounter{equation}{0} \appen{} \label{combi}

The solution of Eq.~(\ref{or}) is presented here.

The equation has the form: \begin{align*} &
\delta_{A,2}[\delta(x_{1}-y_{1})\theta(x_{2}-y_{2})V_{1}(u)]\overleftarrow
{\partial}_{D} \omega^{DB} + 2\omega^{AC} \partial_{C}[V_{2}(z)\delta(x_{1}
-y_{1})\theta(x_{2}-y_{2})]\delta_{B,2}-\\ & \,-\omega^{AC} \partial_{C}
\hat{m}^{1}(z|u)\overleftarrow{\partial}_{D} \omega^{DB} = 0. \end{align*}
Consider all possible combinations of $A$ and $B$:

\subsection { $ {A = i}$, $ {B = j}$} \begin{align*} &
\delta_{i,2}[\delta(x_{1}-y_{1})\theta(x_{2}-y_{2})V_{1}(u)]\overleftarrow
{\partial}_{k} \omega^{kj} + 2\omega^{il} \partial_{l}[V_{2}(z)\delta(x_{1}
-y_{1})\theta(x_{2}-y_{2})]\delta_{j,2}-\\ & \,-\omega^{il} \partial_{l}
\hat{m}^{1}(z|u)\overleftarrow{\partial}_{k} \omega^{kj} = 0. \end{align*}

\subsubsection {$i = j = 1$} $$ \partial_{2}
\hat{m}^{1}(z|u)\overleftarrow{\partial}_{2} = 0\;\Longrightarrow $$
\begin{align*} & \partial_{2}m^{1}(z|u)\overleftarrow{\partial}_{2} =
\sum_{p,q \geq 0} B^{pq}(z,\eta)\partial_{1}^{p} \partial_{2}^{q} \delta(x-y)
= \left( \sum_{p,q \geq 0}B^{\prime pq}(z,\eta)\partial_{1}^{p}
\partial_{2}^{q} \delta(x-y)\right) \overleftarrow{\partial}_{2} + \\ & \, +
\left( \sum_{p \geq 0}[\partial_{2}B^{\prime p}(z,\eta)]\partial_{1} ^{p}
\delta(x_{1}-y_{1})\theta(x_{2}-y_{2})\right) \overleftarrow{\partial }_{2} =
\\ & = \left( \sum_{p,q \geq 0}B^{\prime\prime pq}(u,\eta)\partial_{1}^{p}
\partial_{2}^{q} \delta(x-y)\right) \overleftarrow{\partial}_{2} + \partial
_{2} \left( \sum_{p \geq 0}B_{1}^{p}(z,\eta)\partial_{1}^{p} \delta(x_{1}
-y_{1})\theta(x_{2}-y_{2})\right) \overleftarrow{\partial}_{2}
\;\Longrightarrow \end{align*} \begin{align*} & \partial_{2}m^{1}(z|u) =
\sum_{p,q \geq 0}B^{\prime\prime pq}(u,\eta )\partial_{1}^{p}
\partial_{2}^{q} \delta(x-y) + \partial_{2} \left( \sum_{p \geq
0}B_{1}^{p}(z,\eta)\partial_{1}^{p} \delta(x_{1}-y_{1})\theta(x_{2}
-y_{2})\right) + \\ & \ \ \ \ \ \ \ \ \ \ \ \ \ \ \ \ + \partial_{2}
\mu_{1}(z|y_{1},\eta) = \\ & \, = \partial_{2} \Big( \sum_{p,q \geq
0}B^{\prime\prime\prime pq} (u,\eta)\partial_{1}^{p} \partial_{2}^{q}
\delta(x-y) + \sum_{p \geq 0}[B_{1} ^{p}(z,\eta) +
B_{2}^{p}(\xi,u)]\partial_{1}^{p} \delta(x_{1}-y_{1})\theta (x_{2}-y_{2}) +
\\ & \ \ \ \ \ \ \ \ \ \ \ \ \ \ \ \ + \mu_{1}(z|y_{1},\eta)\Big).
\end{align*} So, \begin{align*} & m^{1}(z|u) = \sum_{p \geq
0}[B_{1}^{p}(z,\eta) + B_{2}^{p}(\xi,u)]\partial_{1} ^{p}
\delta(x_{1}-y_{1})\theta(x_{2}-y_{2}) + \\ & + \mu_{1}(z|y_{1},\eta) +
\mu_{2}(x_{1},\xi|u) + \sum_{p,q \geq 0}B^{\prime \prime\prime
pq}(\xi,u)\partial_{1}^{p} \partial_{2}^{q} \delta(x-y).  \end{align*}

\subsubsection {$i = 1$, $j = 2$} $$ \partial_{2}
\hat{m}^{1}(z|u)\overleftarrow{\partial}_{1} + 2[\partial_{2}
V_{2}(z)]\delta(x_{1}-y_{1})\theta(x_{2}-y_{2})] = 0\;\Longrightarrow $$ $$
\left( \lbrack2\partial_{2}V_{2}(z)]\delta(x_{1}-y_{1})- \sum_{p \geq
0}[\partial_{2}B_{1}^{p}(z)]\partial_{1}^{p + 1} \delta(x_{1}-y_{1})\right)
\theta(x_{2}-y_{2}) + \partial_{2} \hat{\mu}_{1}(z|y_{1},\eta)\overleftarrow
{\partial}_{1} = 0. $$

\subsubsub{} $$ \partial_{2}
\hat{\mu}_{1}(z|y_{1},\eta)\overleftarrow{\partial}_{1} =
0\;\Longrightarrow\partial_{2} \mu_{1}(z|y_{1},\eta)\overleftarrow{\partial
}_{1} = 0\;\Longrightarrow\partial_{2} \mu_{1}(z|y_{1},\eta) = \partial_{2}
\mu _{11}(z|\eta)\;\Longrightarrow $$ $$ \mu_{1}(z|y_{1},\eta) =
\mu_{11}(z|\eta) + \mu_{12}(x_{1},\xi|y_{1},\eta). $$

\subsubsub{} $$ 2[\partial_{2}V_{2}(z)]\delta(x_{1}-y_{1})- \sum_{p \geq
0}[\partial_{2}B_{1} ^{p}(z,\eta)]\partial_{1}^{p + 1} \delta(x_{1}-y_{1}) =
0\;\Longrightarrow\; $$ $$ V_{2} = V_{2}(x_{1},\xi),\;B_{1}^{p} =
B_{1}^{p}(x_{1},\xi,\eta). $$

\subsubsection {$i = 2$, $j = 1$}

$$ \partial_{1} \hat{m}^{1}(z|u)\overleftarrow{\partial}_{2}-[\partial_{2}
V_{1}(u)]\delta(x_{1}-y_{1})\theta(x_{2}-y_{2})] = 0\;\Longrightarrow $$ $$
\left( \sum_{p \geq 0}[\partial_{2}B_{2}^{p}(\xi,u)]\partial_{1}^{p + 1}
\delta(x_{1}-y_{1})-[\partial_{2}V_{1}(u)]\delta(x_{1}-y_{1})\right)
\theta(x_{2}-y_{2}) + \partial_{1} \hat{\mu}_{2}(x_{1},\xi|u)\overleftarrow
{\partial}_{2} = 0. $$

\subsubsub{} $$ \partial_{1}
\hat{\mu}_{2}(x_{1},\xi|u)\overleftarrow{\partial}_{2} =
0\;\Longrightarrow\;\partial_{1} \mu_{2}(x_{1},\xi|u)\overleftarrow{\partial
}_{2} = 0\;\Longrightarrow\mu_{2}(x_{1},\xi|u)\overleftarrow{\partial}_{2} =
\mu_{21}(\xi|u)\overleftarrow{\partial}_{2} \;\Longrightarrow $$ $$ \mu_{2}
= \mu_{21}(\xi|u) + \mu_{21}(x_{1},\xi|y_{1},\eta). $$

\subsubsub{} $$ \sum_{p \geq 0}[\partial_{2}B_{2}^{p}(\xi,u)]\partial_{1}^{p
+ 1} \delta (x_{1}-y_{1})-[\partial_{2}V_{1}(u)]\delta(x_{1}-y_{1}) = 0. $$
So $$ B_{2}^{p} = B_{2}^{p}(\xi,y_{1},\eta),\;V_{1} = V_{1}(y_{1},\eta), $$
$$ \sum_{p \geq 0}[B_{1}^{p}(z,\eta) + B_{2}^{p}(\xi,u)]\partial_{1}^{p}
\delta(x_{1}-y_{1} ) = \sum_{p \geq
0}B_{3}^{p}(y_{1},\xi,\eta)\partial_{1}^{p} \delta(x_{1}-y_{1}).  $$

\subsubsection{$i = 2$, $j = 2$} $$ \{[V_{1}(x_{1},\eta) +
2V_{2}(x_{1},\xi)]\partial_{1} \delta(x_{1}-y_{1} ) +
2[\partial_{1}V_{2}(x_{1},\xi)]\delta(x_{1}-y_{1})\} \theta(x_{2} -y_{2}) +
\partial_{1} \hat{m}^{1}(z|u)\overleftarrow{\partial}_{1} =
0\;\Longrightarrow $$ \begin{align*} & \{2[\partial_{1}V_{2}(x_{1},\xi)] +
[V_{1}(x_{1},\eta) + 2V_{2}(x_{1} ,\xi)]\partial_{1} + \\ & + \sum_{p \geq
0}[B_{3}^{p}(y_{1},\xi,\eta)\partial_{1}^{p + 2} \! + \!  \partial_{1}
B_{3}^{p}(y_{1},\xi,\eta)\partial_{1}^{p + 1}]\} \delta(x_{1}-y_{1})\theta
(x_{2}-y_{2}) \! + \! \partial_{1} \hat{\mu}_{12}^{\prime}(x_{1},\xi|y_{1}
,\eta)\overleftarrow{\partial}_{1} = 0,\\ &
\mu_{12}^{\prime}(x_{1},\xi|y_{1},\eta) = \mu_{12}(x_{1},\xi|y_{1},\eta ) +
\mu_{21}(x_{1},\xi|y_{1},\eta). \end{align*}

\subsubsub{} \begin{align*} \partial_{1}
\hat{\mu}_{12}^{\prime}(x_{1},\xi|y_{1},\eta)\overleftarrow {\partial}_{1} &
= 0\;\Longrightarrow\;\partial_{1} \mu_{12}^{\prime}(x_{1}
,\xi|y_{1},\eta)\overleftarrow{\partial}_{1} = 0\;\Longrightarrow\\
\partial_{1} \mu_{12}^{\prime}(x_{1},\xi|y_{1},\eta) & = \partial_{1} \mu
_{13}(x_{1},\xi|\eta)\;\Longrightarrow\mu_{12}^{\prime}(x_{1},\xi|y_{1}
,\eta) = \mu_{13}(x_{1},\xi|\eta) + \mu_{14}(\xi|y_{1},\eta).\; \end{align*}

\subsubsub{} \begin{align*} & \{[\partial_{1}V_{2}(x_{1},\xi)] +
[2V_{1}(x_{1},\eta) + V_{2}(x_{1} ,\xi)]\partial_{1} + \\ & \, + \sum_{p \geq
0}[B_{3}^{p}(y_{1},\xi,\eta)\partial_{1}^{p + 2} + \partial
_{1}B_{3}^{p}(y_{1},\xi,\eta)\partial_{1}^{p + 1}]\} \delta(x_{1}-y_{1} ) =
0\;\Longrightarrow \end{align*} \begin{align*} B_{3}^{p} & = 0,\;V_{2} =
V_{2}(\xi),\;V_{1}(x_{1},\eta) + 2V_{2}(\xi ) = 0\;\Longrightarrow\\ V_{2} &
= c_{5} = \mathrm{const},\;V_{1} = -2c_{5}. \end{align*} Thus, we obtain that
$m^{1}(z|u)$ can be represented in the form $$ m^{1}(z|u) = \mu_{1}(z|\eta) +
\mu_{2}(\xi|u) + \sum_{p,q \geq 0} B^{\prime\prime\prime
pq}(\xi,u)\partial_{1}^{p} \partial_{2}^{q} \delta(x-y).  $$

\subsection { $A = i$, $B = \beta$} \begin{align*} & \partial_{i}
\hat{m}^{1}(z|u)\overleftarrow{\partial}_{\beta} =
0\;\Longrightarrow\;\partial_{i}
\hat{\mu}_{1}(z|\eta)\overleftarrow{\partial }_{\beta} =
0\;\Longrightarrow\;\partial_{i} \mu_{1}(z|\eta)\overleftarrow
{\partial}_{\beta} = 0\;\Longrightarrow\\ &
\Longrightarrow\;\mu_{1}(z|\eta)\overleftarrow{\partial}_{\beta} = t_{\beta
}(\xi|\eta) = t^{\prime}(\xi|\eta)\overleftarrow{\partial}_{\beta}
\;\Longrightarrow\;\mu_{1}(z|\eta) = t^{\prime}(\xi|\eta) + b_{1}(z).
\end{align*}

\subsection {$A = \alpha$, $B = j$} \begin{align*} & \partial_{\alpha}
\hat{m}^{1}(z|u)\overleftarrow{\partial}_{j} =
0\;\Longrightarrow\;\partial_{\alpha} \hat{\mu}_{2}(\xi|u)\overleftarrow
{\partial}_{j} = 0\;\Longrightarrow\;\partial_{\alpha} \mu_{2}(\xi
|u)\overleftarrow{\partial}_{j} = 0\;\Longrightarrow\\ &
\Longrightarrow\;\partial_{\alpha} \mu_{2}(\xi|u) = p_{\alpha}(\xi |\eta) =
\partial_{\alpha}p(\xi|\eta)\;\Longrightarrow\;\mu_{2}(\xi |u) = p(\xi|\eta)
+ c_{1}(u). \end{align*}

\subsection {$A = \alpha$, $B = \beta$} \begin{align*} \partial_{\alpha}
\hat{m}^{1}(z|u)\overleftarrow{\partial}_{\beta} & =
0\;\Longrightarrow\;\partial_{\alpha}t(\xi|\eta)\overleftarrow{\partial
}_{\beta} = 0,\;t(\xi|\eta) = t^{\prime}(\xi|\eta) +
p(\xi|\eta)\Longrightarrow\\ & \Longrightarrow\;t(\xi|\eta) = b_{2}(\xi) +
c_{2}(\eta). \end{align*}

{}Finally, we obtain $$ m^{1}(z|u) = b(z) + c(u) + \sum_{p,q \geq
0}B^{\prime\prime\prime pq}(\xi ,u)\partial_{1}^{p} \partial_{2}^{q}
\delta(x-y), $$


\end{document}